**Crowdsourcing Star-Formation Research and the Power of Participatory Science**
A Review Article

Grace Wolf-Chase[1], Charles Kerton[2], Kathryn Devine[3], Nicholas Larose[4], & Maya Coleman[5]

[1]Planetary Science Institute
1700 East Fort Lowell, Suite 106
Tucson, AZ 85719, USA
Corresponding Author #1: gwchase@psi.edu
ORCID: 0000-0002-9896-331X

[2]Iowa State University
Department of Physics and Astronomy
2323 Osborn Dr.
Ames, IA 50011, USA
ORCID: 0000-0003-1539-3321

[3]The College of Idaho
2112 Cleveland Blvd.
Caldwell, ID 83605, USA
Corresponding Author #2: KDevine@collegeofidaho.edu
ORCID: 0000-0002-3723-6362

[4]Iowa State University
Department of Physics and Astronomy
2323 Osborn Dr.
Ames, IA 50011, USA
ORCID: 0000-0002-0828-7726

[5]The College of Idaho
2112 Cleveland Blvd.
Caldwell, ID 83605, USA

**Abstract:** We review participatory science programs that have contributed to the understanding of star formation. The Milky Way Project (MWP), one of the earliest participatory science projects launched on the Zooniverse platform, produced the largest catalog of "bubbles" associated with feedback from hot young stars to date, and enabled the identification of a new class of compact star-forming regions (SFRs) known as "yellowballs" (YBs). The analysis of YBs through their infrared colors and catalog cross-matching led to discovering that YBs are compact photodissociation regions generated by intermediate- and high-mass young stellar objects embedded in clumps that range in mass from 10 - $10^4$ $M_\odot$ and luminosity from 10 - $10^6$ $L_\odot$. The MIRION catalog, assembled from 6176 YBs identified by citizen scientists, increases the number of candidate intermediate-mass SFRs by nearly two orders of magnitude. Ongoing work utilizing data from the *Spitzer, Herschel* and *WISE* missions involves analyzing infrared color trends to predict physical properties and ages of YB environments. Methods include applying summary statistics to histograms and color-color plots as well as SED fitting. Students in introductory astronomy classes contribute toward continued efforts refining photometric measurements of YBs while learning fundamental concepts in astronomy through a classroom-based participatory science experience, the PERYSCOPE project. We also describe an initiative that engaged seminaries, family groups, and interfaith communities in a wide variety of science projects on the Zooniverse platform. This initiative produced important guidance on attracting audiences that are underserved, underrepresented, or apprehensive about science.

Keywords: Star Formation; Star Forming Regions; Catalogs; Citizen Science; Educational Software; Milky Way Galaxy; Protoclusters; Protostars

**1. Introduction**

This paper is a response to an invitation by the editor to write an article for *Astrophysics and Space Science* because of my[1] election as a 2024 American Astronomical Society Fellow for "outstanding and sustained work to bring the wonders of astronomical research to the general public, especially to diverse religious communities; and for significant investigations into bipolar molecular outflows within star-forming regions through multi-wavelength observations and analyses." Because bridging research and public outreach has been the central focus of my career, I have elected to write this as a review that focuses on dual aspects of my work in citizen science.[2] One aspect of this work includes star-formation research made possible by a citizen-science discovery that is being performed in collaboration with my co-authors on this paper. Another aspect includes my personal efforts to broaden participation in science through building bridges across cultures and communities that may not otherwise interact. Since my early work on bipolar outflows does not connect directly to my efforts in citizen science, I include it only briefly in personal reflections on my unique career path, which are provided in the Appendix.

The rise of technologies that enable public participation in research alongside professional scientists has made it possible for everyone to contribute to scientific discovery. In this article, we showcase the Milky Way Project (MWP: Milky Way Project 2025) as a wonderful exemplar of participatory science that engaged tens of thousands of volunteers over its lifetime. The MWP illustrates how public participation in science facilitates research that would be impossible or intractable without the input of many people. Furthermore, the MWP demonstrates how serendipitous discoveries made by volunteers can lead to science that was unanticipated by the project's initial design and expectations. Beyond the MWP, I discuss how proactive outreach to engage underrepresented, underserved, and "science-apprehensive"[3] audiences in participatory science can provide a powerful pathway toward diversifying science itself.

---

1 In this paper, the use of "I", "me", and "my" refer specifically to the lead author, Grace Wolf-Chase.

  2 The terms "citizen science" are often replaced with "participatory science" or "participatory research". These terms will be used interchangeably in this article. Public contributions to research include the humanities as well as the sciences.

**1.1 Discovery and Classification in Astronomy**

Classification is a fundamental part of science, and disagreements about how to construct scientific categories often lead to deeper understandings of nature (e.g., Wolf-Chase 2017). Discovery and classification in astronomy go hand-in-hand. Astronomers often begin by classifying astronomical phenomena according to superficial observations that do not reflect the actual nature of the object in question (consider the designation "planetary nebula"). The reader will see numerous examples of this practice throughout this article. Dick (2024a,b) links classification and discovery in astronomy through stages of detection, interpretation and understanding that involve technological developments as well as conceptual and social factors. Historically, initial classifications in astronomy have been based on observable criteria such as color for stars or morphology for galaxies, rather than physical composition, which is not known a priori (Dick 2024a,b). As we learn more about the nature of an object, we revise or amend these categories to reflect our better understanding of its physical characteristics.

Many remote observations are required to determine the nature of astronomical phenomena. These typically include imaging and spectroscopic measurements across much of the electromagnetic spectrum. Hertzsprung-Russell (HR) diagrams were the result of decades of work on the spectral classification of stars. HR diagrams are now powerful tools whereby the nature of diverse types of stars can be inferred from their temperatures and brightnesses. Dick (2024a,b) notes that, unlike stars, distance was not a factor in the classification of galaxies by morphology that resulted in the Hubble Tuning Fork diagram. Galaxy classification based on morphology is still relevant today and has become an activity for citizen scientists participating in *Galaxy Zoo*, which we discuss further below.

Once classifications have been made based on our improved knowledge of an object's nature, new phenomena or objects can often be categorized solely on the basis of directly observable properties. Color-color and color-magnitude diagrams serve as excellent examples of this process. The large number of available archived astronomical surveys provides many opportunities to investigate color trends for different classes of objects. Examples of this include the successful use of *Spitzer Space Telescope* (Werner et al. 2004) infrared (IR) color-color diagrams to identify different categories of young stellar objects (YSOs), identify shocked $H_2$ emission associated with protostellar outflows, and determine physical properties in photodissociation regions (PDRs) (e.g., Ybarra et al. 2014; Kun et al. 2016; Kuhn et al. 2021 and references therein). In the extragalactic realm, Yun et al. (2008) discuss both the merits and limitations of using *Spitzer* IR colors to distinguish between high-redshift submillimetre-bright galaxies and luminous active galactic nuclei (AGNs).

**1.2 Introduction to Participatory Science and Zooniverse**

In recent years, it has become increasingly possible for people of all ages and walks of life to participate in research alongside professionals in many fields of study. Known by various names such as citizen science, community science, and participatory science/research, engaging members of the general public in research enables many major projects that would otherwise be intractable (e.g., Wolf-

---
[3] By this I mean people who, for whatever reasons, are suspicious or fearful of scientific knowledge.

Chase, Hinman, & Trouille 2024; hereafter, WHT24). The project Galaxy Zoo, first launched in 2007, asked volunteers to sort galaxies by their shapes, which enabled the morphological classification of nearly one million galaxies (Lintott et al. 2008). The immense popularity and success of Galaxy Zoo then led to the creation of Zooniverse, the largest and most popular platform for "people-powered research" (University of Oxford News & Events 2019). Since its creation in 2009, Zooniverse has launched more than 500 research projects in many subject areas across the sciences and humanities. At the time of this writing, Zooniverse has amassed nearly 3 million registered volunteers, and Zooniverse research projects have enabled more than 600 academic publications. Over 500 of these publications are in the space sciences (Zooniverse Publications 2025).

In addition to benefiting scientific progress directly, participatory science can empower diverse audiences. This in turn helps improve the general public's perception of science and scientists (Wolf-Chase 2022; WHT24). Research shows that one potential barrier to participation in science is the perception that science and religion have to be in conflict, and that this perception has a stronger negative impact on the participation of women and ethnic minorities (e.g., Barnes and Brownell 2018; Bolger and Ecklund 2022; Ecklund et al. 2019; WHT24). *Engaging Faith-based Communities in Citizen Science through Zooniverse* was an 18-month initiative that was funded by the Alfred P. Sloan Foundation. It was designed to attract and empower audiences that are underserved, underrepresented, or apprehensive about science (Wolf-Chase 2022; WHT24). It leveraged my decades of experience building relationships with diverse religious communities to proactively involve seminaries, youth and family programs, and interfaith organizations in participatory science using Zooniverse projects.

Throughout the duration of the *Engaging* initiative, I led workshops and events to engage faith communities in *Zooniverse* projects on topics that included the space sciences, anthropology, evolution, climate, and ecology (WHT24). The initiative was communicated to approximately 50 organizations via conferences, presentations, newsletters, email, websites, and various types of social media, cumulatively reaching more than 100,000 individuals who may not have been attentive to other venues promoting participatory science (WHT24). Although the pandemic imposed serious limitations, a few common themes emerged during implementation and evaluation of the *Engaging* initiative: people who knew scientists personally were more likely to have positive views of scientists in general; participating in a research project increased participants' confidence in making meaningful contributions to science and increased interest in future participation; and ongoing interactions with professional scientists were extremely important in motivating continuing participation and inspiring confidence in participants (WHT24). The full evaluation report includes results of pre- and post-participation online surveys, focus groups, and informal conversations (Wolf-Chase, Hinman, & Trouille 2021).

One of the most exciting aspects of participatory science is the possibility of discovering something new, something entirely unanticipated by the project's original design. Zooniverse volunteers often identify curious features in images they inspect and bring these features to the attention of the science teams through discussion boards. Occasionally, these features herald important new discoveries. In Galaxy Zoo, volunteers identified "Green Peas," a class of compact extreme star-forming galaxies that were named for their appearance in Sloan Digital Sky Survey (York et al. 2000) images (Cardamone et al. 2009). "Hanny's Voorwerp," an object thought to be a quasar light echo, was named for the Dutch schoolteacher Hanny van Arkel who discovered this unusual object (Lintott et al. 2009). A quick search of these labels using the NASA ADS Abstract service (NASA Astrophysics Data System 2024) reveals that these discoveries have been referenced in well over 100 peer-reviewed scientific articles, with over 50 containing the labels in the article titles.

## 2. A Brief History of the Milky Way Project and Its Products

### 2.1 The Origin of the Milky Way Project

Shortly after the Zooniverse platform was created, Zooniverse's co-founder, Chris Lintott, approached me to discuss the possibility of a project that would advance star-formation studies. I noted that recent large IR surveys conducted by the *Spitzer Space Telescope*, particularly the GLIMPSE (Benjamin et al. 2003; Churchwell et al. 2009) and MIPSGAL (Carey et al. 2009) Legacy surveys of the Galactic Plane, trace various star-formation processes that would be well-suited to visual identifications. Following several months of discussions with Zooniverse team members and scientists within the star-formation community, it was collectively decided that the ninth project to be launched on the Zooniverse platform would utilize the *Spitzer* GLIMPSE and MIPSGAL Legacy surveys to enable volunteers to identify circular mid-IR features (so-called "bubbles") associated with feedback from hot young stars in high-mass SFRs (Benjamin et al. 2003; Churchwell et al. 2006; Churchwell et al. 2007; Churchwell et al. 2009; Carey et al. 2009).

Young, high-mass stars emit copious amounts of UV radiation, which excites polycyclic aromatic hydrocarbon (PAH) molecules and heats dust grains. PAH emission is particularly bright at 8 $\mu$m in PDRs that form bubble rims, and thermal emission from dust grains is often prominent at 24 $\mu$m in bubble interiors (Watson et al. 2008). Understanding high-mass star formation and feedback from high-mass stars is essential for understanding how many stars form, since most stars form in clusters (Lada & Lada 2003; Beuther, Kuiper, & Tafalla 2025, and references therein). Evidence indicates that our Sun formed in an environment similar to the Orion Nebula (Hester et al. 2004), and high-mass stars affect the composition and evolution of planetary systems (e.g., Berné et al. 2024). Furthermore, understanding feedback from high-mass stars is critical to understanding the evolution of galaxies (e.g., Eldridge & Stanway 2022; Thompson & Heckman 2024).

Prior to the MWP, Churchwell et al. (2006, 2007) published bubble catalogs covering the inner Galactic plane ($|l| < 65°$). In total, 591 bubbles were identified by four people who visually inspected GLIMPSE images for 8-$\mu$m ring-shaped features associated with PAH emission from PDRs. "Project 9," which we named the Milky Way Project (MWP), used combined GLIMPSE/MIPSGAL 4.5-$\mu$m, 8-$\mu$m, and 24-$\mu$m images to increase the prominence of bubbles. The MWP enabled volunteers to use an elliptical tool created for the Zooniverse platform to measure the angular sizes and shapes of bubble interiors and bubble rims.

The MWP was launched in December 2010. Within two years, over 10,000 volunteers had identified an order of magnitude more bubbles than in the previous catalogs (Churchwell 2006, 2007). The first MWP data release presented a catalog of positions, radii, thicknesses, and position angles for 5106 IR bubbles across the inner Galaxy (Simpson et al. 2012).

### 2.2 The Discovery of "Yellowballs"

Within months of its launch, volunteers began using #yellowball in the MWP's discussion board to draw the project team's attention to compact, roundish features that appeared yellow in the r-g-b (24 - 8 - 4.5 $\mu$m) color scheme that was used for the *Spitzer* images (see Figure 1).

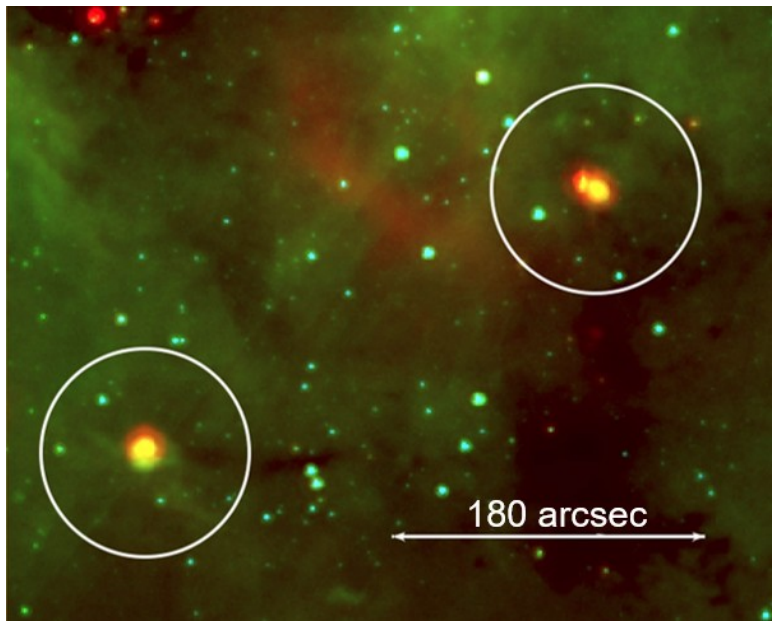

**Figure 1. Yellowballs (YBs).** Typical YBs are compact and appear yellow in *Spitzer* 4.5 (blue)-8.0 (green)-24 (red) $\mu$m mid-IR images used by the MWP.

Volunteers tagged these so-called "yellowballs" (YBs) on a casual basis, which resulted in 928 identifications (Kerton et al. 2015; hereafter, KWA15). Through a combination of catalog cross-matching and IR color analysis, KWA15 showed that YBs highlight a mix of compact star-forming regions (SFRs), including ultra-compact and compact HII regions, as well as analogous regions for less massive B-type stars. The majority of these YBs were unambiguously associated with dense molecular clumps identified in the ATLASGAL catalog of compact 870 $\mu$m sources (Csengeri et al. 2014) or the Bolocam Galactic Plane Survey (BGPS: Aguirre et al. 2011; Dunham et al. 2011; Ginsburg et al. 2013) catalog. The distinctive color of YBs in the MWP images arises from the cospatial emission by PAHs (bright at 8 $\mu$m) and warm dust (bright at 24 $\mu$m), which is expected for compact, thick PDRs prior to the formation of extended bubbles (KWA15).

This interpretation of YBs as compact PDRs was supported by the finding that the average $F_{12\mu m}/F_{8\mu m}$ (F12/F8) flux ratio of a representative sample[4] of 183 YBs was lower than expected for HII regions or planetary nebulae (PNe) (Anderson et al. 2012). The strength of PAH emission features around 7-8 $\mu$m is highly sensitive to the PAH ionization state, such that a more compact PDR will have a higher PAH ionization fraction and stronger emission around 7-8 $\mu$m (Roelfsema et al. 1996; Draine & Li 2007). Draine & Li (2007) showed that the strength of the lines around 8 $\mu$m change by factors of order 10 when going from neutral to ionized PAHs. KWA15 noted that many YBs do not appear in the Red *MSX* Survey (RMS: Lumsden et al. 2013) catalog. The RMS catalog is considered to be the largest statistically selected catalog of high-mass YSOs (MYSOs) and HII regions, complete to the detection of a B0 V star at the distance of the Galactic center. However, it only includes objects with rising spectral energy distributions (SEDs) through the mid- and far-IR (Lumsden et al. 2013). As MYSOs develop thick PDRs, they become brighter at 8 $\mu$m than 12 $\mu$m (KWA15). KWA15 concluded that YBs represent a new class of objects transitioning from the earliest stage of protostellar or protocluster evolution to more evolved bubbles. They suggested that, just as studies of "Green Peas" have yielded critical insights into galaxy evolution (Amorín et al. 2012; Chakraborti et al. 2012), further study of YBs could yield critical insights into the evolution of protostars and protoclusters.

---

[4] See KWA15 for a discussion of how this sample was chosen.

The human-interest nature of the discovery of YBs led to international media attention[5] and an invitation to the authors to write the first space science article for *Frontiers for Young Minds*, an open-access science journal[6] whose articles are written by scientists and reviewed by kids (Wolf-Chase & Kerton 2015). Many groups have since been engaged in the story of how YBs were discovered and what we are learning from them, including a minority-serving intergenerational summer science camp co-organized by Hood Theological Seminary and the Families and Communities Together organization (Wolf-Chase 2022; WHT24).

**2.3 The Continuing Impact of the Milky Way Project**

The MWP was relaunched in September 2016 with a new interface that included the identification of YBs, candidate bow-shock driving stars, and bubbles as principal goals. Cygnus-X Complex (~ 24 square degrees centered on $\ell$~79.3°, $b$~1°: Beerer et al. 2010; Hora et al. 2007) and *Spitzer* Mapping of the Outer Galaxy (SMOG: 21 square degrees, $\ell$=102° to 109°, $b$=0° to 3°: Carey et al. 2008) images were added to images of the inner Galactic Plane ($|\ell|$<65°: Benjamin et al. 2003; Churchwell et al. 2009; Carey et al. 2009) to enable identifications in diverse Galactic SFRs. The bubble-drawing tool was updated for improved bubble shape and size measurements, and images were presented at a maximum zoom level that enabled users to identify smaller bubbles (Jayasinghe et al. 2019). The reliability of bubble identifications was assessed by comparison to the DR1 catalog, which enabled more rigorous elimination of spurious features (Jayasinghe et al. 2019). The second MWP data release produced a final catalog of 2600 bubbles, ~30% fewer than the DR1 catalog, as well as a catalog of 599 bow shocks produced by high-mass stars (Jayasinghe et al. 2019). Furthermore, the positions and angular sizes of 6176 YBs were obtained by aggregating volunteer identifications across the roughly 300 square degrees covered by the MWP (Wolf-Chase et al. 2021; hereafter WKD21).

The MWP has enabled critical studies of the effects of high-mass star formation across the inner Galactic Plane. Kendrew et al. (2012) performed a statistical study which indicated that roughly 22% of high-mass young stars may have formed as a result of feedback from expanding HII regions. Beaumont et al. (2104) demonstrated how machine learning (ML) combined with crowdsourced training data from bubbles identified by citizen scientists can address the weakness of using each approach alone. Kendrew et al. (2016) published a statistical study that showed that temperatures and column densities of cold, dense clumps near bubbles were significantly higher than those in the field. Furthermore, they showed that high-mass star formation was enhanced in the vicinity of bubbles, with the highest increase near the largest bubbles.

The MWP has also enabled work by independent teams of scientists. For example, Xu & Offner (2017) utilized the MWP bubble catalog to evaluate the capabilities of *Brut*, a ML algorithm, at identifying bubbles using synthetic dust observations. A recent publication by Nishimoto et al. (2025) used MWP bubbles to train a deep-learning algorithm that can accurately detect both Galactic and extragalactic (LMC and NGC 628) bubbles using *Spitzer* 8-$\mu$m and 24-$\mu$m data and *James Webb Space Telescope* (*JWST*) 7.7-$\mu$m data.

---

[5] e.g., three initial press releases; 1453 NASA retweets and 1953 favorites by Feb. 5, 2015; Jan. 31, 2015 Astronomy Picture Of the Day (APOD); several radio interviews; and over a dozen popular articles and press release reprints, including international media outlets in Europe and Asia as well as the U.S.
[6] This journal has been named a great website for kids by the American Library Association.

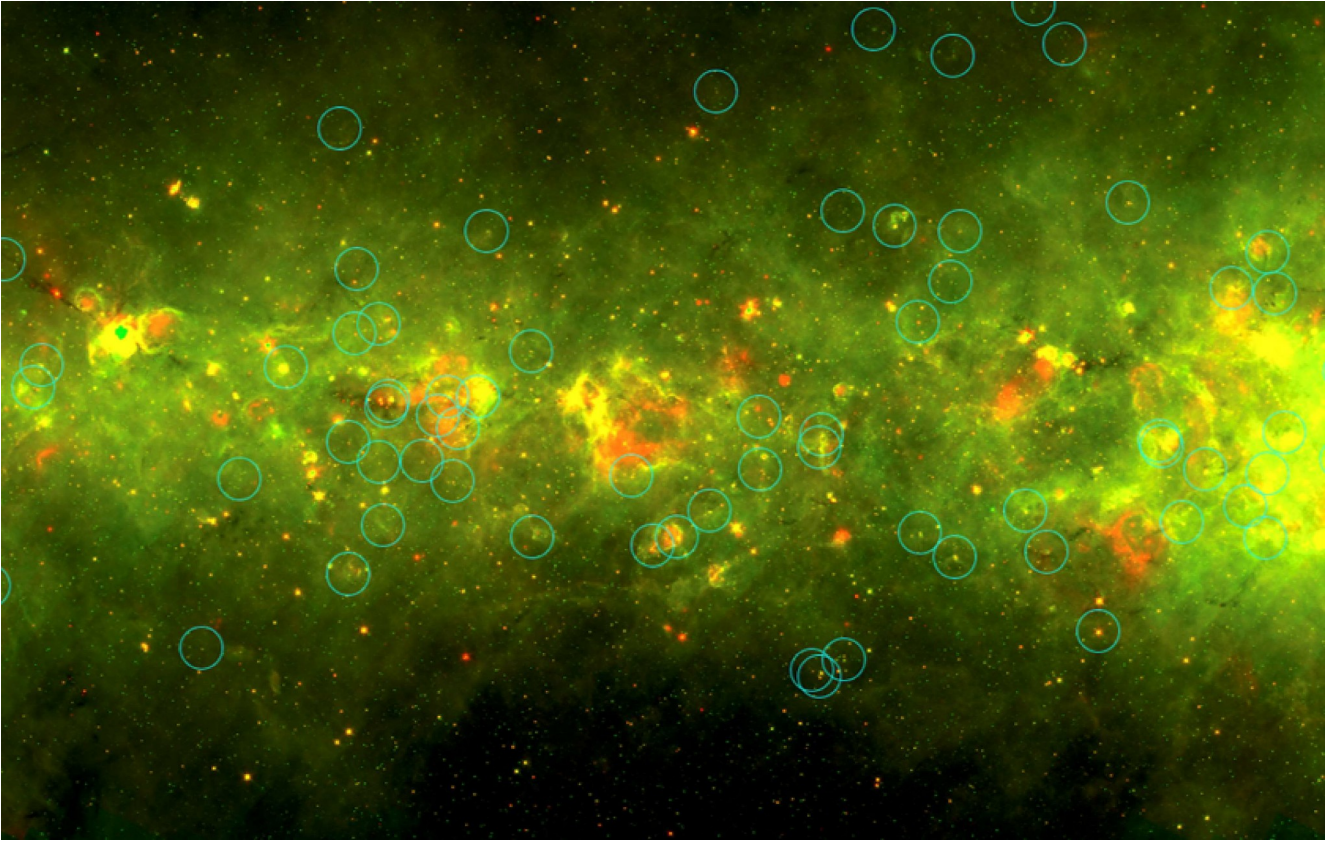

**Figure 2. New Star-Forming Regions.** Yellowballs (YBs) with associations in the Hi-GAL Compact Source Catalogue (Elia et al. 2017) that lack counterparts in catalogs of HII regions and high-mass young stellar objects (MYSOs) in the 6-square-degree region of the Milky Way between $\ell$ = 31.5°-34.5°, $b$ = ± 1°. This two-color *Spitzer* image displays 8-$\mu$m emission (green) and 24-$\mu$m emission (red).

Explicitly including YBs as objects of interest in the 2016 launch of the MWP enabled volunteers to identify and measure the angular sizes of 6176 YBs. This resulted in the identification of many lower-luminosity YBs that were missed in the earlier identifications (WKD21). Figure 2 displays YBs in a 6-square-degree region of the Milky Way that lack counterparts in catalogs of HII regions and MYSOs. Section 3 presents evidence that as many as 70% of all YBs trace intermediate-mass (IM)SFRs where the most massive stars are in the 3-8 $M_\odot$ range. Below 3 $M_\odot$, stars would not produce PDRs generating the 8-$\mu$m emission seen in YBs (Dewdney et al. 1991; Diaz-Miller et al. 1998; Kerton 2002). Thanks to the efforts of citizen scientists, the YB catalog (presented in Section 4) increases the number of candidate IMSFRs by nearly two orders of magnitude!

## 3. Bridging Low- and High-Mass Star Formation with "Yellowballs"

Astrophysical research has identified two clear regimes of star formation: low-mass ($\leq$ 1-2 $M_\odot$) and high-mass ($\geq$ 10 $M_\odot$). The first regime, in which only low-mass stars are formed, involves the collapse of molecular cores whose mass distribution mimics the stellar initial mass function (e.g., André 2017; Motte et al. 2018). These stars form in loose, low (stellar) density aggregates, and the molecular cores can effectively be considered in isolation from the surrounding interstellar medium (ISM). In contrast, the second regime includes the formation of both high- and low-mass stars in high (stellar) density clusters. The high-mass stars form from multi-scale accretion onto low-mass stellar embryos (e.g., the various scenarios outlined by Motte et al. 2018), or from the collapse of massive, turbulently supported cores (e.g., McKee & Tan 2003; Krumholz & McKee 2005).

IMSFRs, where the most massive stars are in the 3-8 $M_\odot$ range, bridge low- and high-mass star-formation regimes. In their recent review, Beuther et al. (2025) argue that we need to move toward a unified picture of star formation that accounts for the similarities as well as qualitative and quantitative differences between low- and high-mass star formation. Critical questions they outline include: How do processes in isolated low-mass SFRs like Taurus vary in comparison with the intense feedback processes from MYSOs? How important are different feedback processes in cluster formation? Understanding intermediate-mass star formation may be key to understanding the transition between low- and high-mass star formation. However, the study of IMSFRs has been severely hampered because, due to their low IR luminosity, the number of these regions that had been previously identified was relatively small (e.g., Arvidsson et al. 2010; Lundquist et al. 2014, 2015). As mentioned at the end of Section 2.3, YBs present an ideal identifier with which to expand the number of studied IMSFRs.

**3.1 Physical Properties and Evolutionary Stages of YBs**

WKD21 conducted a pilot study of 516 YBs (~8% of the entire YB database) in a 20-square-degree region between Galactic coordinates $\ell$ = 30°-40° and $b$ = ± 1°. The researchers chose to examine this region because it overlaps with many star-formation catalogs and surveys, including the *Herschel* Hi-GAL (Molinari et al. 2010) Compact Source Catalogue (CSC: Elia et al. 2017); the GaussClump Source Catalogue from ATLASGAL (Csengeri et al. 2014); the Co-Ordinated Radio "N" Infrared Survey for High-mass star formation catalog (CORNISH: Purcell et al. 2013); the RMS catalog (Lumsden et al. 2013); the *WISE* Catalog of Galactic HII Regions (Anderson et al. 2014); and the Boston University – Five College Radio Astronomy Observatory Galactic Ring Survey (BU–GRS: Jackson et al. 2006). Furthermore, a sample of 516 YBs provided a tractable amount of data with which the team could conduct a more robust color analysis than the KWA15 study.

The WKD21 study supported the earlier KWA15 finding that the majority of YBs are compact PDRs, which are typically sub-parsec in size. Approximately 75% (368) of YBs in the WKD21 sample are associated with dense star-forming cores (diameter D < 0.2 pc) or clumps (0.2 $\leq$ D $\leq$ 3 pc) (Bergin & Tafalla 2007) identified in the *Herschel* Hi-GAL (Molinari et al. 2010) CSC (Elia et al. 2017). Elia et al. (2017) built SEDs for Hi-GAL clumps using fluxes obtained from associations with *WISE* (Wright et al. 2010), *MSX* (Lumsden et al. 2013), MIPSGAL (Carey et al. 2009), *Herschel* PACS and SPIRE (Poglitsch et al. 2010; Griffin et al. 2010), ATLASGAL (Csengeri et al. 2014), and BGPS (Aguirre et al. 2011; Ginsburg et al. 2013) sources. Where possible, they used fluxes at 21 *μ*m, 22 *μ*m, 24 *μ*m, 70 *μ*m, 160 *μ*m, 250 *μ*m, 350 *μ*m, 500 *μ*m, 870 *μ*m, and 1100 *μ*m to construct SEDs. To explore the physical properties of YB-associated Hi-GAL clumps, WKD21 used distances they calculated independently to rescale clump properties reported by Elia et al. (2017). The WKD21 procedure for determining distances to YBs is outlined briefly at the beginning of Section 4.

Figure 3 shows bolometric luminosity versus clump mass plots for YB-associated clumps. The position of a clump in the plots is related to its evolutionary stage. Each plot shows model evolutionary tracks for protostellar clumps (red solid lines), and each track corresponds to a clump with a given initial envelope mass (Molinari et al. 2008). YBs with (upper panels: RMS and WISE Type K) and without (lower panels: WISE Type Q and YB/Hi-GAL Only) known MYSO and HII-region counterparts are shown. Some YBs have both RMS and WISE Type K counterparts. These are shown on both panels. The vertical tracks correspond to when embedded protostars are rapidly gaining mass from their birth cloud. The luminosity of the cloud increases due to accretion and internal heating, while the mass of the clump is lowered reflecting the conversion of gas into stars. The horizontal tracks correspond to the slower process of cloud dispersal.

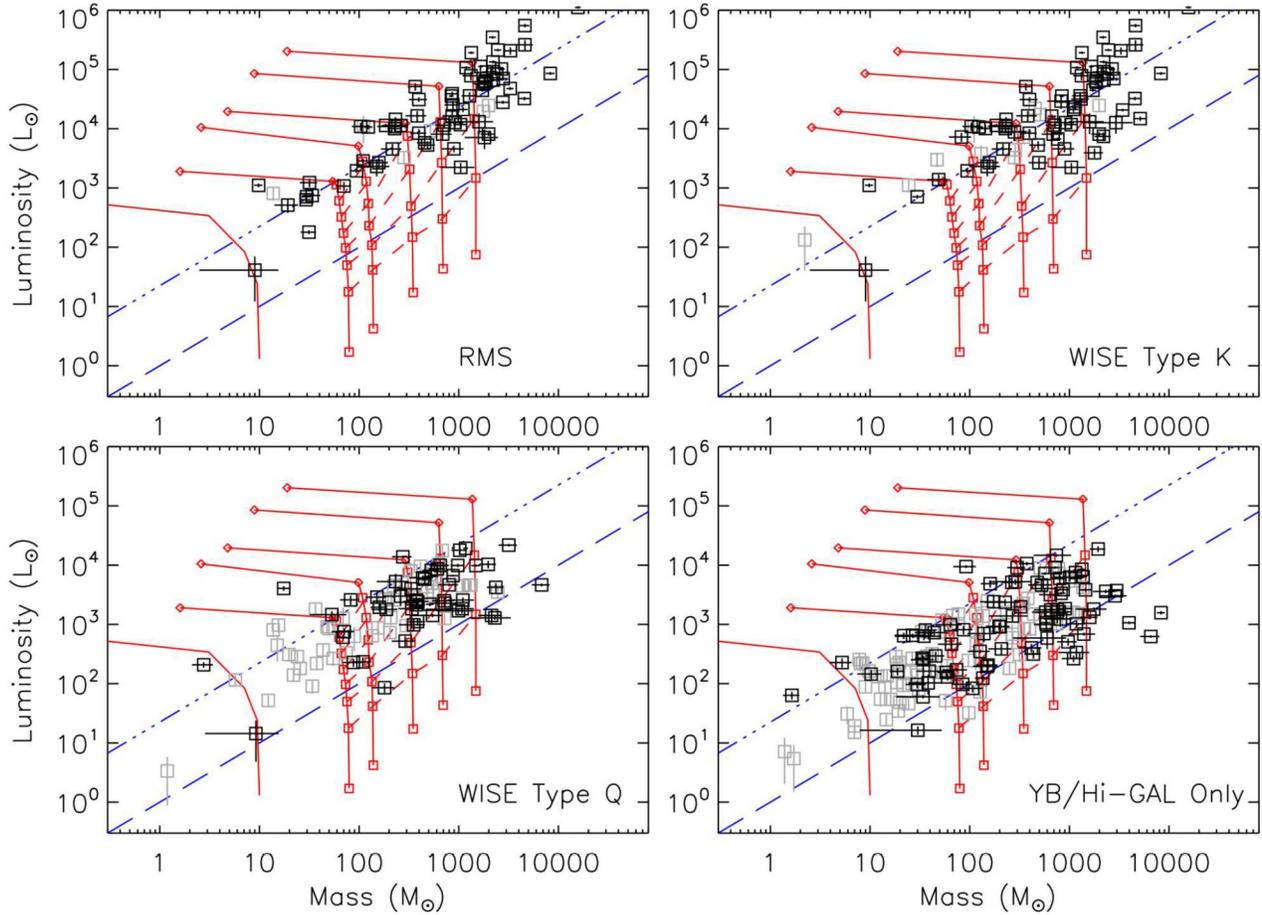

**Figure 3. Bolometric Luminosity vs. Clump Mass for YB-clump matched subsets** (adapted from WKD21, Figure 4). YBs with (upper panels) and without (lower panels) known HII-region and MYSO counterparts are shown. Red lines: clump evolutionary tracks (Molinari et al. 2008). Small red squares: $5 \times 10^4$ year intervals. Red dashed lines: isochrones connecting different clump masses. Blue diagonal lines: 90th percentile lower limit for HII regions (upper line) and 90th percentile upper limit for pre-stellar objects (lower lines) (Elia et al. 2017). Light symbols are sources with large mass uncertainty (mass error bars not shown).

Although the evolutionary tracks shown in Figure 3 were based on the SED models for the evolution of individual protostellar cores (Molinari et al. 2008), subsequent models reinforce the robustness of the luminosity vs. mass diagram as a diagnostic tool for the evolutionary classification of dense star-forming clumps (Molinari et al. 2019). Figure 3 highlights differences between clumps that have associations with known HII regions and MYSOs and those without. While most WISE Type K and RMS sources lie near the horizontal-track transition, nearly all WISE Type Q and YB/Hi-GAL Only sources fall along the vertical tracks expected for actively accreting protostars and protoclusters with M < $10^3$ $M_\odot$ and L < $10^4$ $L_\odot$.

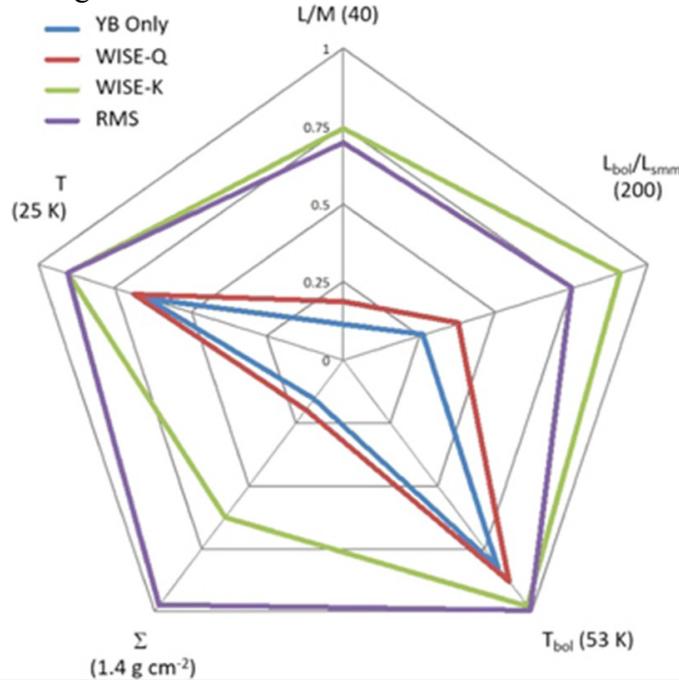

**Figure 4. YB Distance-independent Quantities** (adapted from WKD21, Figure 5). The radar plot shows five distance-independent quantities (described in the main text) for four different classes of YB-associated Hi-GAL clumps. L/M has units of $L_\circ/M_\circ$. YBs associated with HII regions and MYSOs occupy distinctly different regions in this parameter space. Values plotted are medians, and the maxima are indicated for each radial line. The numbers along the vertical radial line indicate the fraction of the maximum value for each quantity. Different classes are connected via different colored pentagons. RMS and WISE-K subsets are almost entirely high-mass SFRs, while the WISE-Q and YB Only subsets are not.

Figure 4 shows a "radar plot" of distance-independent quantities related to protostellar/protocluster evolutionary stage and mass for YB/Hi-GAL matched clumps. Clump properties are taken from Elia et al. (2017). L/M ($L_\circ/M_\circ$) is the bolometric luminosity to clump mass ratio. As can be seen by inspecting the evolutionary tracks in Figure 3, this ratio increases as protostars/protoclusters evolve. $L_{bol}/L_{smm}$ is the ratio of bolometric luminosity to luminosity in the submillimeter ($\lambda$ = 350 $\mu$m - 1100 $\mu$m). It can be used as a rough proxy for the evolution of star-formation activity: as more stars form within a clump, an increasing proportion of the luminosity shifts to shorter wavelengths. The temperature of a blackbody that has the same mean frequency as the observed SED ($T_{bol}$: Myers & Ladd 1993) increases as star-formation activity heats the clump. T is the temperature derived from greybody fits to *Herschel* SEDs at $\lambda \geq 160$ $\mu$m, which also increases as protostars/protoclusters evolve. The gas surface density ($\Sigma$) is often used in conjunction with theoretical models to identify high-mass SFRs (e.g., Krumholz & McKee 2008; López-Sepulcre et al. 2010). The median of each of the five quantities for the four subsets depicted in Figure 3 is plotted along one of five axes. Each axis is normalized to the value indicated at the end of the axis to account for the different numerical ranges spanned by each quantity.

Figure 4 shows a clear difference in the median values of distance-independent quantities for each of the four subsets of YB/Hi-GAL matched clumps displayed in Figure 3. YBs associated with RMS and WISE-K sources have significantly higher $L_\odot/M_\odot$, $L_{bol}/L_{smm}$, and $\Sigma$ median values than YB Only and WISE-Q-associated YBs. These differences can be attributed to both evolutionary stage and stellar mass content. WKD21 concluded that YB-Only and WISE-Q subsets likely contain a mix of IMSFRs and very young high-mass SFRs. The low median $L_\odot/M_\odot$ and $\Sigma$ values are expected for IMSFRs. Specifically, the median $\Sigma$ for these subsets fall below predicted thresholds for high-mass star formation (e.g., Krumholz & McKee 2008; López-Sepulcre et al. 2010). However, low $L_{bol}/L_{smm}$ and $T_{bol}$ values are also consistent with very young high-mass SFRs. We discuss the use of IR colors to further discriminate between age and stellar content in Sections 3.2 and 4.1.

In summary, WKD21 demonstrated that the majority of YBs are associated with Hi-GAL clumps (Elia et al. 2017) that range in mass from 10 - $10^4$ $M_\odot$ and luminosity from 10 - $10^6$ $L_\odot$. Only ~ 30% of YBs in the WKD21 study are associated with HII regions and MYSOs. Although some of the protoclusters highlighted by YBs may contain protostellar embryos destined to become high-mass stars, distance-independent properties support the conclusion that a majority of YBs highlight IMSFRs.

**3.2 The Infrared Colors of YBs**

Using data from the InfraRed Astronomical Satellite (IRAS: Neugebauer et al. 1984), Kerton (2002) showed that IMSFRs have IR colors that are distinct from high-mass SFRs. Specifically, IMSFRs have lower $F_{25\mu m}/F_{12\mu m}$ and higher $F_{60\mu m}/F_{25\mu m}$ flux ratios compared with high-mass SFRs. This difference can be understood in terms of the combined effect of a lowering of the dust temperature (increasing the $F_{60\mu m}/F_{25\mu m}$ ratio), and the increased survivability of PAH dust grains in less harsh UV environments expected within IMSFRs (Giard et al. 1994), which acts to lower the $F_{25\mu m}/F_{12\mu m}$ ratio (Kerton 2002). These early results suggest that measuring IR colors of YBs using contemporary facilities (e.g., *Spitzer*, *WISE*, *Herschel*) could provide a fruitful way forward in distinguishing between high-mass SFRs and IMSFRs.

Examining color trends for YBs requires placing stringent constraints on the accuracy of YB flux measurements. This is challenging, as YBs are often extended sources that are located in complex, structured IR backgrounds associated with the Galactic Plane. As a consequence, traditional aperture and PSF-fitting photometry are not effective techniques for measuring their fluxes. To circumvent this problem, WKD21 developed a semi-interactive Python-based tool[7] that they used to conduct photometry on YBs at 8.0 μm (GLIMPSE), 12 μm (*WISE)*, and 24 μm (MIPSGAL).[8] This tool enables the user to interactively select points surrounding a YB to create a mask. The code subsequently interpolates over the area within those points using a multiquadric radial basis function to create a background estimate. The difference between background and image is used to create a source-only image from which the flux density is calculated at each wavelength.

The primary source of uncertainty in this method is users' differing point selections, which is a long-recognized issue for extended source photometry at IR wavelengths (e.g., Fich & Terebey 1996). This uncertainty can be minimized and quantified by having multiple users take measurements of each source. The WKD21 study minimized photometric errors by using six independent measurements per source. Photometry was performed by the research team and their students. Photometric results were only reported for sources with standard deviation/mean flux <0.5, which was judged as a reasonable cut-off on the relative error. Figure 5 presents photometry results for YBs in the WKD21 sample that have Hi-GAL clump counterparts (Elia et al. 2017). Hi-GAL 70-$\mu$m fluxes were used for this sample, and only YBs meeting the standard deviation/mean flux <0.5 photometric cut-off at 8 $\mu$m, 12 $\mu$m, and 24 $\mu$m are plotted.

---

[7] https://github.com/astrodevine/YB_Photometry
[8] After WKD21 was published, the tool was modified to include performing 70-μm photometry using *Herschel* data (Molinari et al. 2010).

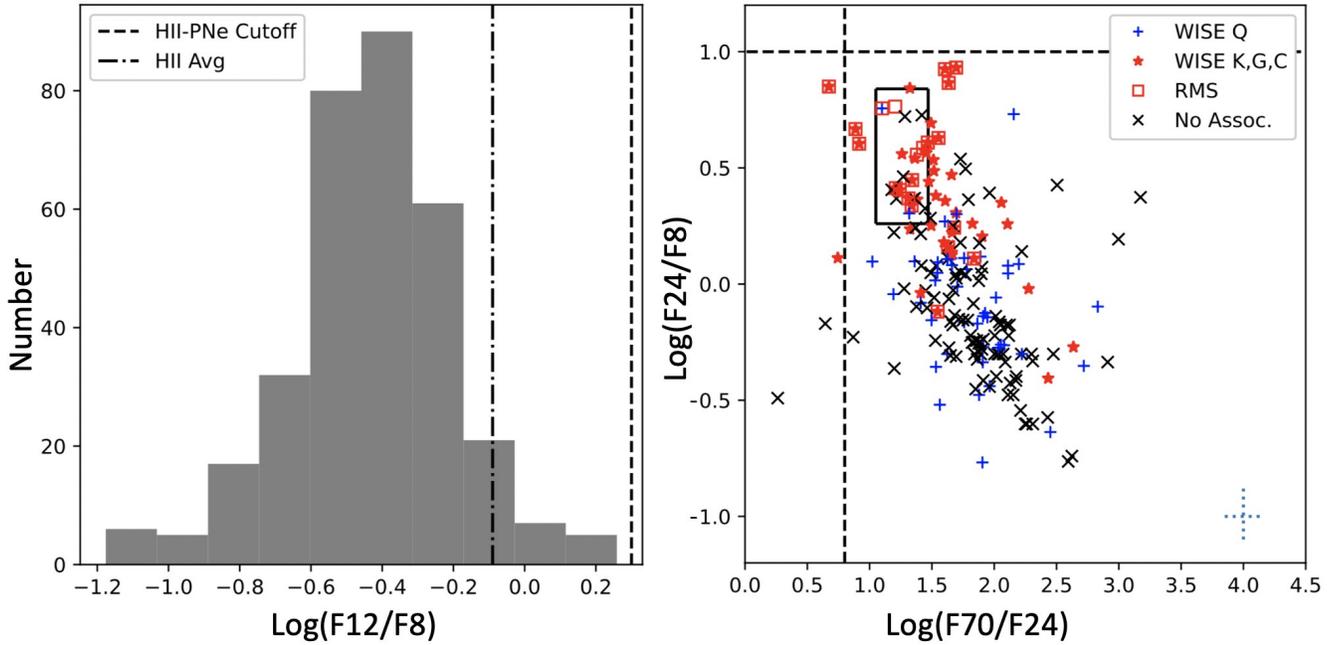

**Figure 5. YB IR Colors** (adapted from WKD21, Figures 6 & 7). The majority of YBs have IR colors that are distinct from the colors of HII regions and MYSOs. Left: Distribution of log(F12/F8) colors for 324 YBs with Hi-GAL clump associations. Dot-dashed line: average color of HII regions; dashed line: color separation of HII regions and Planetary Nebulae (PNe) (Anderson et al. 2012). Right: Color-color plot of 219 YBs with Hi-GAL clump associations. Colors and shapes indicate YBs with counterparts in other catalogs. Solid rectangle: region of mean colors of HII regions. Vertical and horizontal dashed lines: IR color cut-offs for distinguishing HII regions from PNe (Anderson et al. 2012). Dotted lines in the lower right show average color uncertainty.

The mean log(F12/F8) color of the YBs in the histogram in the left panel of Figure 5 is -0.43±0.15 (WKD21), significantly lower than the average color of *WISE* HII regions (-0.09±0.11; Anderson et al. 2012). On average, YB PDRs are more compact than typical HII regions, thus causing the color shift to lower log(F12/F8) colors. The right panel of Figure 5 displays a color-color plot of log($F_{24\mu m}/F_{8\mu m}$) vs. log($F_{70\mu m}/F_{24\mu m}$) flux ratios (log(F24/F8) and log(F70/F24), respectively). The WISE K,G,C label refers to known, grouped, and candidate HII regions identified by Anderson et al. (2014), respectively. YBs associated with these objects, as well as HII regions and MYSOs identified in the RMS (Lumsden et al. 2013) survey (shown in red), are grouped near the expected colors for HII regions. YBs without such associations (shown in blue and black) are predominantly located in a separate region of the plot corresponding to lower log(F24/F8) and higher log(F70/F24) colors. This color shift would be expected for lower-mass star-forming clumps due to the combined effect of the decrease in dust temperature and the increased survivability of the 8-$\mu$m emitting PAHs in the less harsh ultraviolet (UV) radiation environment (Giard et al. 1994; Kerton 2002; Draine & Li 2007; Watson et al. 2008).

**3.3 Yellowballs in the Broader Context of Star Formation**

Sections 3.1 & 3.2 demonstrate that most YBs trace accreting protoclusters that bridge the low- and high-mass regimes. Since YBs offer "snapshots" into the development of protoclusters across a broad range of physical properties and ages (Figures 3 & 4), their further study can facilitate answering the question: what causes the transition from low-mass (isolated) to high-mass (clustered) star formation? The ability to answer this question has important implications for planetary science as well as astronomy. The composition and evolution of planetary systems that develop in the company of high-mass stars are likely to differ from those that do not (e.g., Berné et al. 2024); therefore, understanding these different modes of star formation may be critical to understanding the diversity of exoplanets.

While the WKD21 study provided intriguing evidence that color trends are associated with both physical properties and evolutionary stages of YB environments, the sample size was insufficient to draw significant conclusions about these relationships. In Section 4, we introduce a catalog that includes all 6176 YBs. Roughly 4000 of these YBs have clump/core associations (Elia et al. 2017, 2021; Cao et al. 2019) for which it is possible to derive physical properties, thus extending the earlier study by more than an order of magnitude. The new catalog provides a robust sample of objects to explore how IR colors vary with both distance-dependent quantities such as mass and luminosity, and distance-independent quantities associated with stellar content and evolutionary stages of SFRs.

**4. The MIRION Catalog of Yellowballs and Its Applications**

All 6176 YBs identified across the ~300-square-degree region surveyed by the MWP have been compiled into the Mid-InfraRed Interstellar Objects and Nebulae (MIRION[9]) Catalog of Yellowballs (Devine et al. 2025, in prep). This catalog reports: fluxes at 8.0 $\mu$m, 12 $\mu$m, 24 $\mu$m, and 70 $\mu$m; flux errors based on five measurements per source per wavelength; flagging based on visual inspection; distance determinations; and YB cross-matches with other catalogs (Devine et al. 2025, in prep). The designation "yellowballs" is an example of problematic nomenclature that pervades labels used in astronomy. It derives from the appearance of these objects in the specific representative color scheme (r-g-b: 24 - 8.0 - 4.5 $\mu$m) used for the MWP. The name of the catalog reflects the fact that YBs were identified by their unique mid-IR appearance and acknowledges the fact that while the vast majority of YBs identify young SFRs, the catalog includes other compact objects that are bright at mid-IR wavelengths. This is reflected in cross-matches and flagging results that indicate that not all MIRION sources are sites of star formation (Devine et al. 2025, in prep).

Distances to YBs in the MIRION catalog were determined by using a Bayesian calculator (WKD21; Devine et al. 2025, in prep). The calculator uses Galactic coordinates (l, b) and radial velocity as input parameters, and considers kinematic distance, displacement from the Galactic Plane, and proximity to individual parallax sources to determine a distance probability density function for each source (Reid et al. 2016, 2019). Where possible, the calculator utilizes ancillary information, such as associations with masers, that can help break the ambiguity inherent in determining distances using radial velocities toward the inner Galaxy (e.g., Reid et al. 2019; Mège et al. 2021). Radial velocities of molecular clouds toward YBs were determined by examining $^{13}$CO spectra toward YBs from CO surveys of the Galactic Plane. Where multiple strong $^{13}$CO velocity components were identified (in roughly 15% of cases), emission lines from dense gas tracers such as NH$_3$ (e.g., Wienen et al. 2012; Hogge et al. 2018) or OH surveys that trace star formation (e.g., Beuther et al. 2016) were used to identify the YB-associated clumps (Devine et al. 2025, in prep).

---

9 For Tolkien fans, we note that the word "Mirion" refers to the Silmarils (jewels) in Sindarin Elvish. (See https://www.elfdict.com/w/mirion).

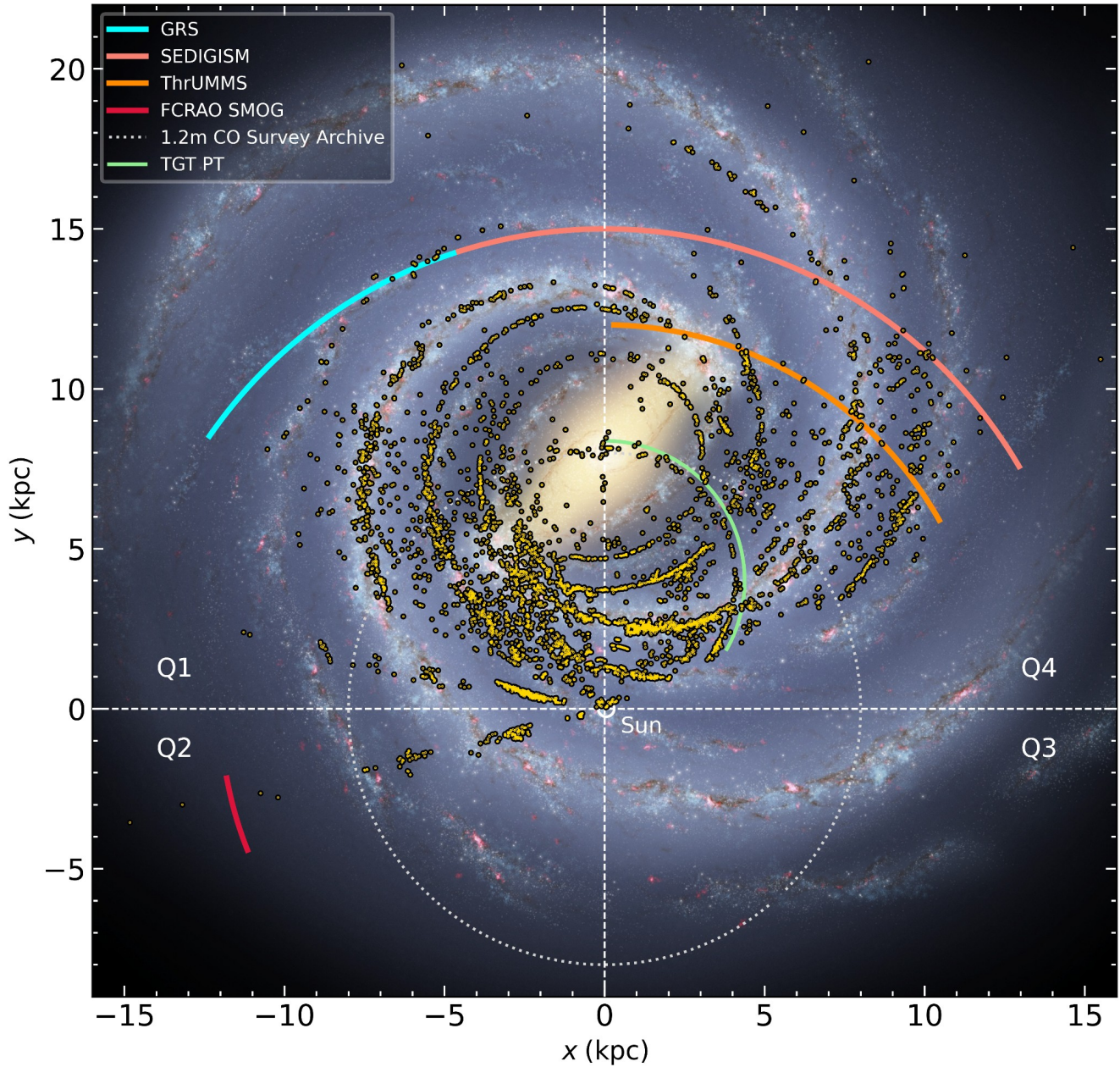

**Figure 6. YBs and CO Surveys.** This top-down map of the Milky Way illustrates the Galactic longitude distribution of the $^{12}$CO and $^{13}$CO surveys used to determine distances to YB-associated molecular clouds and the location of all 6176 YBs (yellow-filled circles). Q1 - Q4 refer to the Galactic quadrants. Colored arcs indicate longitude coverage of surveys. The green arc "TGT PT" highlights a locus of YBs where the measured velocities toward these sources exceed the maximum allowed for the rotation model at a given ($\ell$, $b$). In such cases the YB distances were set to the tangent-point distance. This accounts for <3% of all YBs. (GRS: Jackson et al. 2006; SEDIGISM: Schuller et al. 2021; ThrUMMS: Barnes et al. 2015; FCRAO SMOG refers to CGPS: Taylor et al. 2003; 1.2-m CO Survey Archive: Dame et al. 2001).

Figure 6 illustrates the CO surveys that were used for determining distances to YB-associated molecular clouds and the Galactic distribution of all 6176 YBs identified by MWP participants. The surveys include the Boston University Galactic Ring Survey (GRS: Jackson et al. 2006); the Structure, Excitation and Dynamics of the Inner Galactic Interstellar Medium survey (SEDIGISM: Schuller et al. 2021); the Three-mm Ultimate Mopra Milky Way Survey (ThrUMMS: Barnes et al. 2015); the Canadian Galactic Plane Survey (CGPS: Taylor et al. 2003); and the 1.2-m CO Survey Archive (Dame et al. 2001). Distances to 94% of the YBs were determined using the method discussed above, while non-kinematic distances were adopted for the remaining YBs (Devine et al. 2025, in prep).

The MIRION Catalog will be an invaluable resource that can be used to select objects by distance, mass, luminosity, age, surface density, etc., for detailed investigations into the relationship between IR colors and properties of MIRION sources. Roughly 300 YBs lie at distances < 2 kpc (Devine et al. 2025, in prep). These relatively nearby SFRs would make excellent targets for high-resolution studies with facilities such as the Atacama Large Millimeter/ submillimeter Array (ALMA), Very Large Array (VLA), and *JWST*. For example, such studies could involve investigating differences in formation mechanisms at play across low- and high-mass SFRs by determining the spatial distribution of protostellar cores (e.g., Motte et al. 2018) and measuring the velocity structure at different spatial scales, from large-scale flows to accretion onto individual protostellar embryos.

Sections 4.1 and 4.2 take a brief look at some of the current and future applications of the new catalog to enhance our understanding of star formation across different mass/luminosity regimes and evolutionary stages. The MIRION catalog will be hosted online after initial publication and updated periodically to incorporate future photometry performed by introductory astronomy students who participate in the project described in Section 4.3 below. Crowdsourcing these measurements will provide continual refinement of fluxes and flux uncertainties.

**4.1 How are IR Colors Related to the Properties and Ages of Star-Forming Regions?**

The MIRION Catalog (Devine et al. 2025, in prep) expands the work of WKD21, providing a richer data set from which to associate IR colors with physical properties. Figure 7 presents color-color plots similar to the one in the right panel of Figure 5, but for hundreds, rather than a few tens, of sources in each cross-matched category. Due to the large number of sources in each image, the images show the concentration of sources located in a color space, with darker color indicating a higher number of sources. The RMS and WISE C,G,K sources peak within the boxed color space associated with HII regions, as expected. In contrast, YBs with WISE Q and no associations trace distinctly different populations. Their peaks are clearly located outside of the color space expected for HII regions, consistent with results found by WKD21 (See Section 3.2).

YBs with WISE Q and no associations have significantly lower mean (± s.d.) $\log(F24/F8)$ colors (0.12±0.47 and 0.04±0.44, respectively) compared to YBs with WISE C, G, K and RMS associations (0.50±0.43 and 0.71±0.34, respectively). This suggests that most YBs with WISE Q and no associations are compact PDRs surrounding intermediate-mass stars or very young high-mass stars (Giard et al. 1994; Roelfsema et al. 1996; Kerton 2002; Draine & Li 2007; KWA15; WKD21). However, the distinction between $\log(F70/F24)$ colors of the different groups is unclear (WISE Q: 1.36±0.39; WISE C,G,K: 1.22±0.38; RMS: 1.00±0.47; No Associations: 1.01±0.60). This result will be explored further in future work.

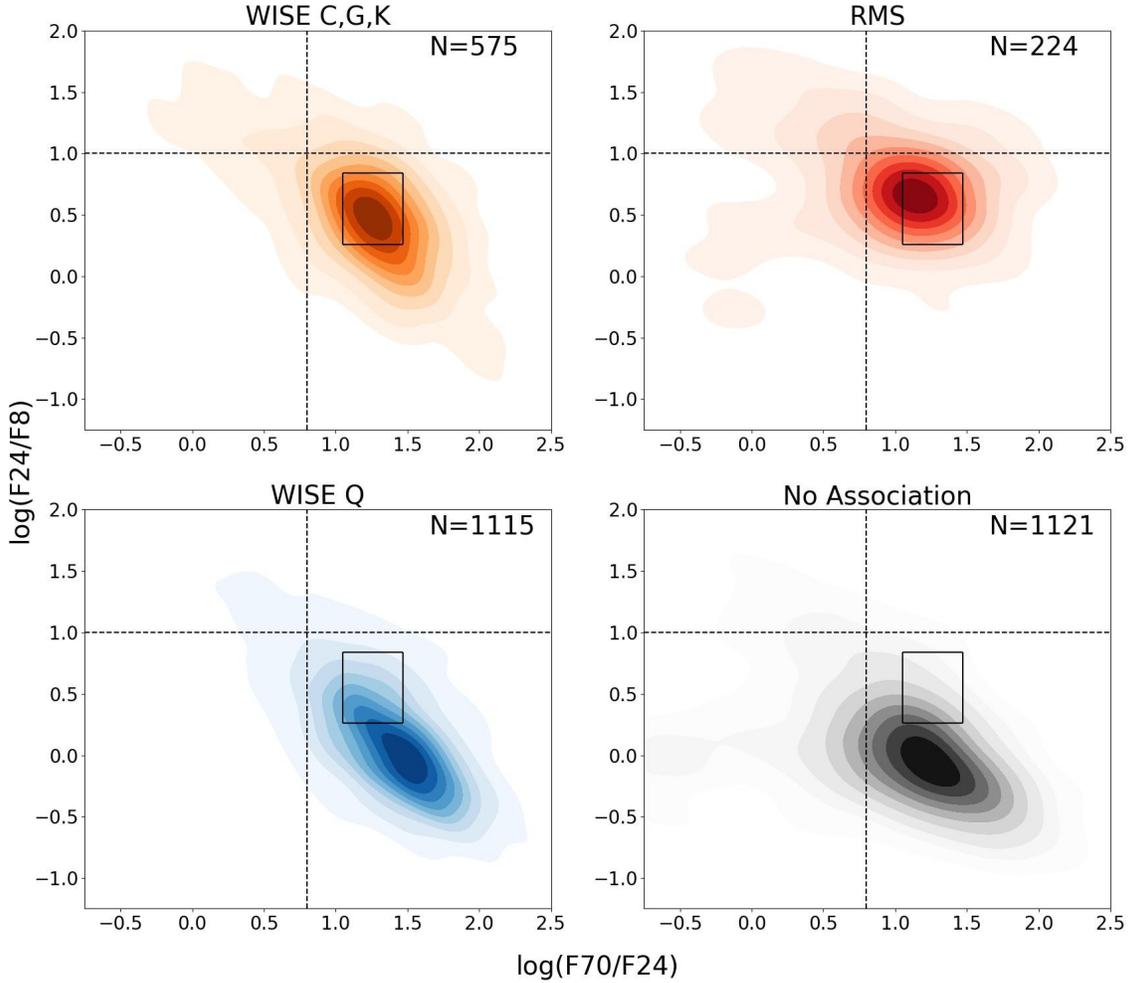

**Figure 7. YB IR Colors** (adapted from Devine et al. 2025, in prep). Color-color plots of YBs, sorted by associations with other catalogs. Top Left: Plot of 575 YBs with WISE C,G,K HII region associations (mean log(F24/F8) = 0.50±0.43, mean log(F70/F24) = 1.22±0.38). Top Right: Plot of 224 YBs with RMS source associations (mean log(F24/F8) = 0.71±0.34, mean log(F70/F24) = 1.00±0.47). Bottom Left: Plot of 1115 YBs with WISE Q associations (mean log(F24/F8) = 0.12±0.43, mean log(F70/F24) = 1.36±0.39). Bottom Right: Plot of 1121 YBs with no associations with HII regions or MYSOs (mean log(F24/F8) = 0.04±0.44, mean log(F70/F24) = 1.03±0.60). As in Figure 5, the solid rectangle indicates the region of mean colors of HII regions, while vertical and horizontal dashed lines are the IR color cut-offs for distinguishing HII regions from PNe (Anderson et al. 2012).

This expanded sample of flux measurements can be associated with physical properties for roughly 4000 YBs associated with clumps/cores (Elia et al. 2017, 2021; Cao et al. 2019). Figure 8 presents histograms of the luminosity and mass distributions of *Herschel* CSC clumps and cores associated with YBs. The median luminosity and mass of 784 clumps associated with WISE C,G,K HII regions are $1.1 \times 10^4$ L$_\odot$ and 590 M$_\odot$, while the median luminosity and mass of 3159 clumps with no or WISE Q associations are 898 L$_\odot$ and 177 M$_\odot$, respectively. Two-sample Kolmogorov-Smirnov (KS) tests showed that there is a highly statistically significant difference between the mass and luminosity distribution of each group: D(784, 3159) = 0.26, p<.001 and D(784, 3159) = 0.45, p<.001, respectively. These results are fully consistent with the WKD21 study and expand the number of candidate IMSFRs by nearly two orders of magnitude.

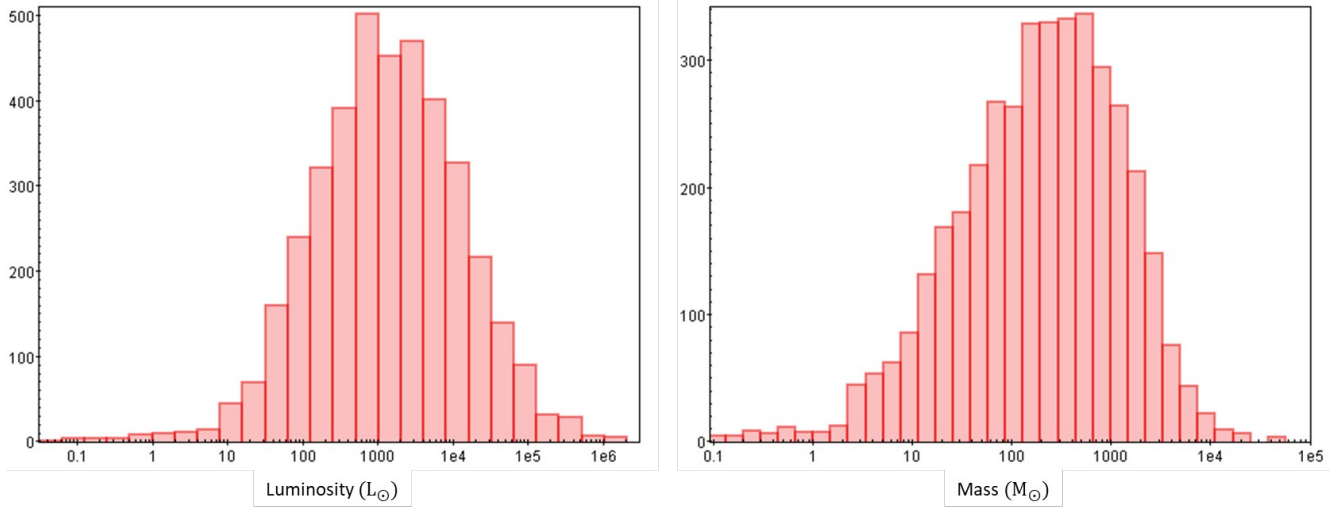

**Figure 8. Physical Properties of *Herschel* Hi-GAL clumps/cores hosting YBs.** Histograms of the luminosity (left) and mass (right) of *Herschel* CSC clumps and cores associated with YBs. Properties were initially derived by Elia et al. (2021) and have been rescaled using newly determined distances (Devine et al. 2025, in prep).

Photometry results at 8 *μ*m, 12 *μ*m, 24 *μ*m, & 70 *μ*m are being used to examine color trends utilizing several different techniques. Where possible, these wavelengths are being supplemented with longer-wavelength fluxes from the Hi-GAL CSC catalog (Elia et al. 2017, 2021). These techniques include applying summary statistics to histograms and color-color plots as well as fitting SEDs using available libraries (Robitaille 2017; Zhang & Tan 2018). Furthermore, ML regression approaches (e.g., Partial Least-Squares: Dyar et al. 2012; Dyar et al. 2016; LASSO: Hastie et al. 2009; Jolliffe 2002; Wegelin 2000) are being employed to predict distance-independent numerical variables $L_\odot/M_\odot$, $L_{bol}/L_{smm}$, $\Sigma$, $T$, and $T_{bol}$ from IR colors (Wolf-Chase et al. 2025, in prep).

Figure 9 presents one example of how a color-color plot can elucidate evolutionary differences for a sample of YBs that have associated clump masses in the *Herschel* Hi-GAL (Molinari et al. 2010) CSC (Elia et al. 2017, 2021). By selecting a particular mass range, it is possible to isolate the relationship between source luminosity and IR colors. In this example, we see a diagonal color trend for YBs associated with clump masses greater than 700 $M_\odot$. This mass range was chosen as a sample of clumps that in the Molinari et al. (2008) model (see Figure 3) represent high-mass SFRs at different stages of evolution. The color trend is associated with a change from younger and less luminous objects in the lower right of the plot, to more evolved and more luminous objects in the upper right. Astrophysically, this trend is caused by the increasing temperature of the dust grains in YB environments over time.

Future research will focus on more detailed investigations of the relationship between YB colors and their environments. For example, YB populations can be isolated by age to investigate the difference between mass and IR colors. Sources could be binned by timesteps along vertical protostellar tracks and by mass along isochrones in luminosity vs. mass diagrams (see Figure 3) to determine the relationship between YB color, mass, and age. Histograms can also be used to uniquely identify subsets of YBs (e.g., grouped by age, luminosity, association with HII regions), similar to the way that IR colors can be used to separate HII regions from PNe (Anderson et al. 2012). The mean F24/F8 colors of YBs that are and are not associated with HII regions and MYSOs shown in Figure 7 are significantly different. The striking difference in the color likely reflects the increasing hardness of the UV radiation field with increasing stellar mass, which decreases the strength of the 8-$\mu$m emission (Giard et al. 1994; Kerton 2002).

The distance-independent quantities ($L_\circ/M_\circ$, $L_{bol}/L_{smm}$, $\Sigma$, T, and $T_{bol}$) shown in the radar plot (Figure 4) are related to evolutionary stage and mass. There are three widely-used theoretical thresholds for identifying environments capable of producing high-mass stars: $\Sigma > 1$ g cm$^{-2}$ (Krumholz & McKee 2008), $\Sigma > 0.3$ g cm$^{-2}$ (López-Sepulcre et al. 2010), and $M(r) > 870 M_\odot (r/pc)^{1.33}$, where r is the radius within which mass is measured (Kauffmann & Pillai 2010). Color distinctions or shifts across these different thresholds may better distinguish the transition between low- and high-mass SFRs.

It is likely that molecular clumps are not isolated regions of star formation but rather accrete gas while simultaneously forming stars. Several models have been proposed for mass accretion onto molecular clumps. For example, in the conveyor-belt model of cluster formation, gas continues to be accreted onto a clump while star formation is occurring (Longmore et al. 2014; Krumholz, McKee, & Bland-Hawthorn 2019). Similarly, hierarchical fragmentation envisions collapse at various scales within larger structures (Vázquez-Semadeni et al. 2009, 2019). The exact nature of clump accretion and how it depends on the mass of the clump is model-dependent. However, there is theoretical support for cloud-to-clump accretion (e.g., Svoboda et al. 2016), and direct observational evidence for large-scale flows onto clumps and filaments (Kumar et al. 2020; Liu et al. 2023).

Competing effects of accretion and mass loss due to stellar feedback in protoclusters ensure that distinguishing SFRs that will eventually produce high-mass stars from those that will not is a non-trivial problem. For example, the surface density of star-forming clumps can increase as they accrete more material and decrease when outflows and HII regions disrupt their natal environments. However, observed IR color shifts as the stellar content of protoclusters evolve (e.g., Figure 9) suggest a promising approach; investigating correlations between $\Sigma$, color, and age could help observationally distinguish between IMSFRs and very young high-mass SFRs on a probabilistic basis (Larose et al. 2025, in prep).

**4.2 Understanding Formation Mechanisms Across Different Star-Forming Regimes**

Roughly 300 YBs in the MIRION Catalog lie at distances <2 kpc. At 2 kpc, a resolution of 2″, easily achievable with ALMA and the VLA, corresponds to 0.01 pc, which is comparable to predicted sizes of stellar embryos in massive protoclusters (Motte et al. 2018). Therefore, the MIRION Catalog will provide an ample supply of relatively nearby targets to study formation mechanisms at play across low- and high-mass SFRs by measuring the velocity structure at different spatial scales, from large-scale hub-filament flows (e.g., Kumar et al. 2020; Wells at al. 2024) to accretion flows at the individual core/stellar embryo scale (Motte et al. 2018).

Accretion onto protostars is universally accompanied by outflow activity (e.g., Beuther et al. 2025, and references therein). Outflows exhibit great diversity in morphology – most are bipolar, although some display wide-angle winds, while highly-collimated jets are also observed across the YSO mass spectrum (e.g., Caratti o Garatti et al. 2015). A few outflows are nearly isotropic and some display explosive behavior. An excellent review of the observations that elucidate outflow diversity and evolution, and the importance of outflows in regulating star formation, can be found in Bally (2016). Follow-up observations of the closest MIRION catalog sources will provide valuable observational data studying the interplay between accretion and outflow across a wide range of star-forming environments.

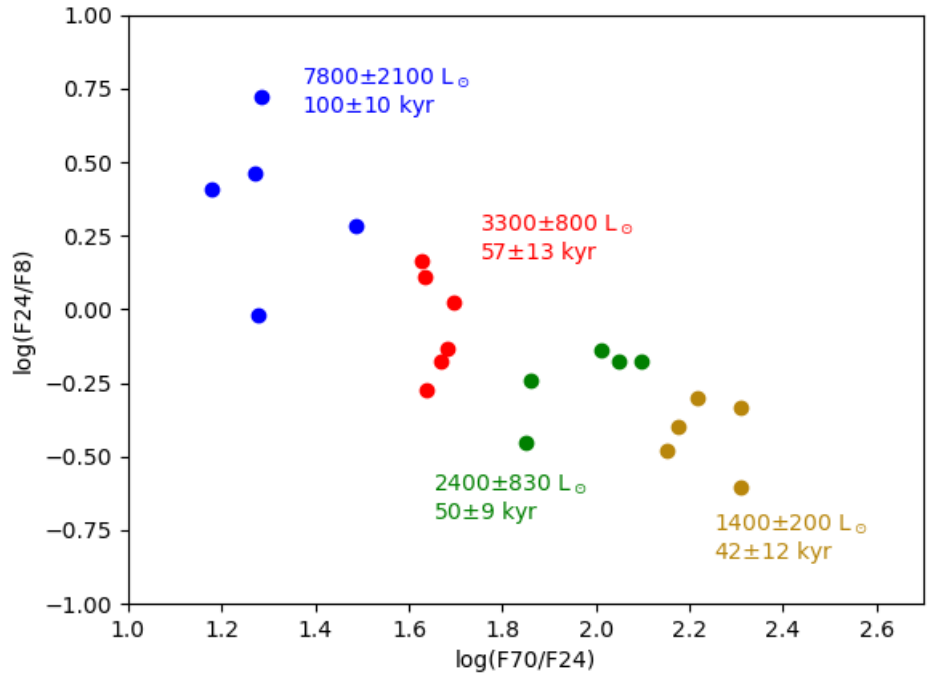

**Figure 9. Luminosity/Age and IR Color.** IR colors for a sample of YBs associated with *Herschel* compact sources with M > 700 $M_\odot$ are plotted. Average color uncertainty is ~0.2 dex (for clarity error bars are not shown). Quartiles in the log(F70/F24) color are designated by blue, red, green, and gold colors. The average luminosity and age of the YBs in each quartile are shown beside each subsample. The color distribution is clearly related to the change from younger/less luminous objects in the lower right to older/more luminous objects in the upper left.

For example, high-resolution follow-up studies will be useful in probing complicated, compact star-forming environments. Most stars form in compact clusters or small groups of stars (Lada & Lada 2003), and protoclusters exhibit multiple outflows. Wolf-Chase, Arvidsson, & Smutko (2017) reported a search for near-IR Molecular Hydrogen emission-line Objects (MHOs) associated with low-resolution (~30″) CO outflows towards 26 regions containing high-mass protostellar candidates (Molinari et al. 1996, 1998, 2000, 2002; Zhang et al. 2005). The SFR Mol 11 is an excellent example of how complicated protocluster evolution is at the ~1″ resolution of the near-IR images. This region displays two distinct epochs of star formation (Figure 10). Near the center of the frame is a PDR surrounding a young cluster visible at near-IR wavelengths, while to the northeast lies a younger cluster of protostars that is not visible at near-IR wavelengths. These protostars are embedded in seven dense cores (Lee et al. 2011), and they drive jets delineated by a series of MHOs along several distinct position angles.

In some high-mass SFRs, collimated outflows from YSOs might be punctuated by short-duration violent eruptive phenomena generated by interactions between protostars; however, due to the large distances to most high-mass SFRs, only a few "explosive" outflows, such as the Orion BN/KL and DR21 outflows, have been studied in great detail (Bally 2016; Bally et al. 2017; Bally et al. 2020; Bally 2024; Guzmán Ccolque et al. 2024; Wright et al. 2024). How common are "explosive" outflows? Do they occur exclusively in the most high-mass SFRs? Why do some protoclusters exhibit jets emanating in all directions (e.g., Mol 11) while others show alignments with dense filaments (e.g., Kong et al. 2019)? With the *JWST*, it is now possible to resolve details of outflows from protoclusters even in the extreme outer Galaxy (Izumi et al. 2024). Understanding the cumulative impact of outflows in protocluster environments is essential to understanding star formation rates, efficiencies, and the end products of star formation.

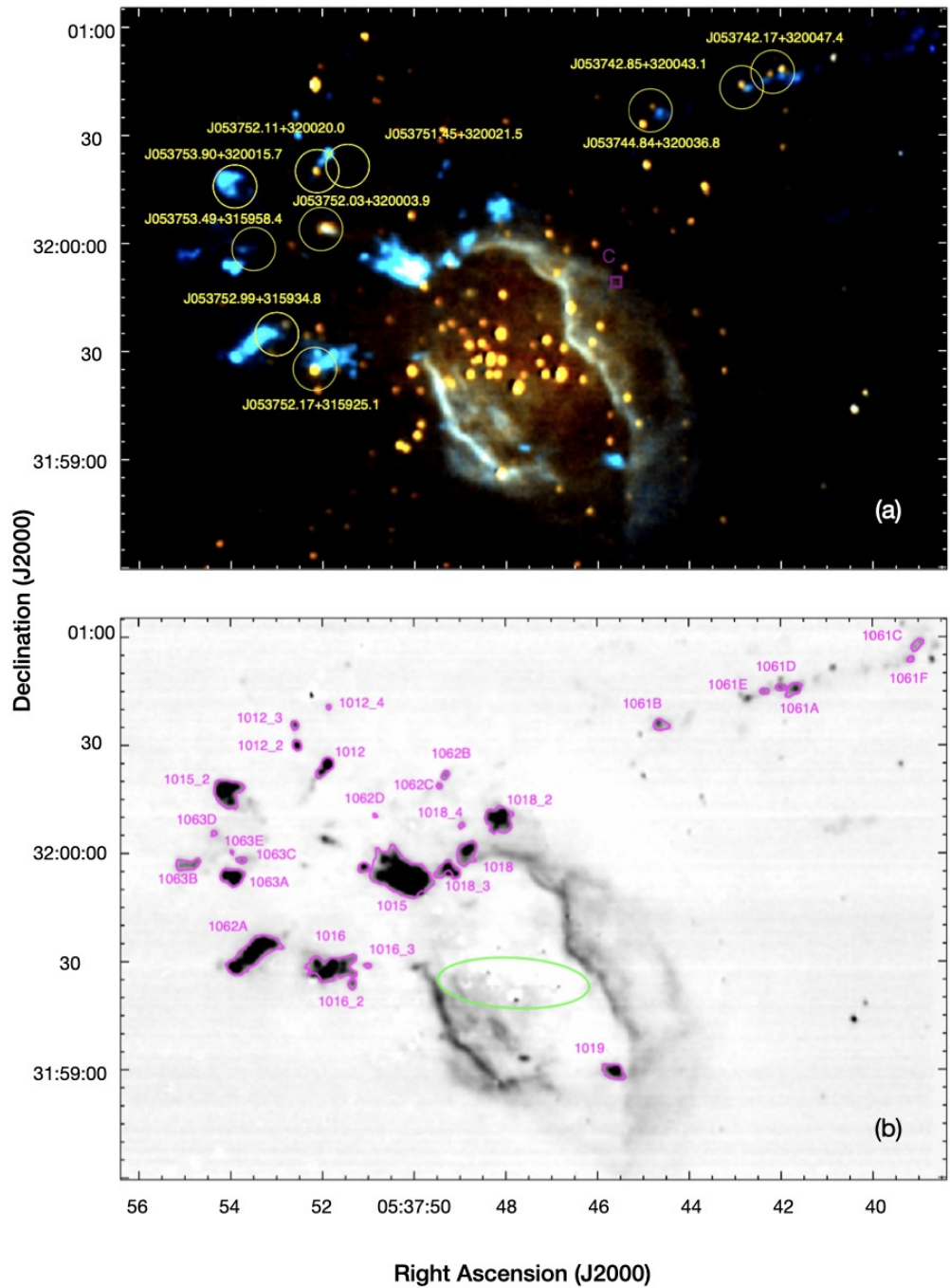

**Figure 10. Two Distinct Epochs of Star Formation in Mol 11** (adapted from Wolf-Chase, Arvidsson, & Smutko 2017, Figure 6). (a) r-g-b image combining continuum-subtracted $H_2$ 2.12 μm line emission (blue), continuum-subtracted $H_2$ 2.25 μm line emission (red), and average of the continuum-subtracted $H_2$ 2.12 μm and $H_2$ 2.25 μm images (green). ALLWISE (Wright et al. 2010; Mainzer et al. 2011) sources that fit Class 0/I protostellar color criteria (yellow circles) and the position of a candidate HII region (purple box labelled "c") in the WISE Catalog of Galactic HII Regions (Anderson et al. 2014) are indicated. (b) continuum-subtracted $H_2$ 2.12 μm grayscale image with MHO numbers and regions (magenta) and *IRAS* source error ellipse (green) indicated.

Protostellar outflows have been identified in the vicinity of many YBs. More than 300 YBs are associated with outflows detected by the SEDIGISM survey (Yang et al. 2022). Furthermore, roughly 20% of entries in the Extended Green Objects (EGOs) catalogs that were produced from the *Spitzer* Legacy GLIMPSE I & II surveys of the Galactic Plane (Cyganowski et al. 2008; Chen et al. 2013) have associations with YBs in the MIRION Catalog. EGOs are bright 4.5-$\mu$m features that were identified from their visual appearance in GLIMPSE images that use an 8.0 - 4.5 - 3.6 $\mu$m red-green-blue (r-g-b) color scheme. Their prominence at 4.5 $\mu$m is due to emission from shocked $H_2$ in outflows from MYSOs (Cyganowski et al. 2008; Chen et al. 2013).

As discussed throughout this paper, IR colors can be extremely useful in distinguishing properties associated with different astronomical phenomena. Ybarra et al. (2014) developed color diagnostics to separate shocked $H_2$ emission associated with protostellar outflows from UV-excited emission generated by PDRs using *Spitzer* 3.6-µm, 4.5-µm, & 5.8-µm data. The application of these color diagnostics to the MIRION catalog could form the basis for a statistical study of outflows associated with a range of protogroup/cluster masses and evolutionary stages. Such a study could shed light on the frequency and timescales of outflow events in different SFRs and help identify the best targets for detailed studies of outflows and their driving sources with *JWST* and ALMA observations. Comparison with $H_2$ 2.12-$\mu$m images from the United Kingdom Infrared Telescope Wide Field Infrared Survey for $H_2$, an unbiased survey of the Galactic Plane from $\ell \sim 357°$-$65°$ and $b = \pm 1.5°$ (Froebrich et al. 2015), could further facilitate identifying protostellar jets (e.g., Chauhan et al. 2024).

**4.3 Participatory Science via the PERYSCOPE Project**

The MIRION catalog contains thousands of YBs that can be used to explore the extent to which IR colors can predict the properties and ages of SFRs. Multiple measurements are required in order to investigate the accuracy of measured fluxes and constrain errors, as well as identify and flag unreliable sources. The People Enabling Research: a Yellowball Survey of the Colors of Protostellar Environments (PERYSCOPE) Project (PERYSCOPE 2024) was designed to increase the number of YB flux measurements by enabling introductory astronomy students to assist with the photometry data collection process.

Students participating in PERYSCOPE measure YB fluxes using a simplified, web-based version of the photometry code used by the research team (Section 3.2). The code, which is implemented on the Google CoLab platform, enables students to plot and analyze their results using histograms and color-color diagrams. It is accompanied by a classroom activity that guides students to draw conclusions from their plots and connects their efforts to fundamental concepts and tools widely used in astronomy, such as blackbody radiation and HR Diagrams. The full activity can be completed in approximately three hours, consisting of pre-class preparation, in-class photometry, and post-class reflection components. Students only need a computer and internet access to use the web-based tool. Following completion of the activity, students choose whether they want to donate their work to the MIRION Catalog and be acknowledged on the online site and in any future publications.

PERYSCOPE received one of the first NASA Citizen Science Seed Funding Program (CSSFP) grant awards (Talbot 2021) to develop an astronomy laboratory exercise that engages students in actual research. Participation in PERYSCOPE involves the employment of important scientific practices such as data collection, analysis, and formulating follow-up questions; the discovery of an unknown outcome (i.e., what is the distribution of YB colors and how do colors correlate with YB properties?); broad relevance to our understanding of star formation; collaboration within the classroom and with the science team; and iteration (revisiting and flagging outlier sources, adapting in response to student findings).

Although non-major introductory astronomy college students are the primary target audience, PERYSCOPE can also be used in astronomy classes at the high-school level. Astronomy instructors can express their interest in using the code and classroom activities through the PERYSCOPE site.[10] PERYSCOPE illustrates how participatory science continues to be interwoven into the study of YBs. An online version of the MIRION catalog will be updated periodically to incorporate photometry performed by introductory astronomy students who participate in PERYSCOPE. Crowdsourcing these measurements will provide continual refinement of fluxes and flux uncertainties.

## 5. Summary

### 5.1 Crowdsourcing Star Formation Research

Participatory research has contributed to the advancement of science in many different fields. In this article, we have focused on volunteers' contributions to star-formation research. The MWP achieved its initial goal of producing the largest catalog to-date of "bubbles" associated with feedback from hot young stars in high-mass SFRs (Simpson et al. 2012; Jayasinghe et al. 2019). Furthermore, it has led to subsequent research on the effects of feedback from high-mass stars on their birth environments (e.g., Kendrew et al. 2012; Kendrew et al. 2016) and studies illustrating the effectiveness of using volunteer identifications to improve ML techniques (e.g., Beaumont et al. 2104; Xu & Offner 2017; Nishimoto et al. 2024).

An unanticipated result of the MWP was the identification of over 6000 so-called "yellowballs" (YBs) by MWP volunteers. Significantly, the MIRION catalog increases the number of candidate IMSFRs by nearly two orders of magnitude. YBs provide "snapshots" into the evolution of star-forming clumps ranging in mass from $10 - 10^4 \, M_\odot$ and luminosity from $10 - 10^6 \, L_\odot$ (KWA15; WKD21). They present a unique opportunity to study protoclusters forming high-mass and intermediate-mass stars throughout the accretion phase (Section 3.1). YBs show IR color trends that are linked to clump properties such as mass, luminosity, gas surface density, and other quantities associated with the stellar content and evolutionary stages of developing protoclusters (Sections 3.2 & 4.1). Participants in the PERYSCOPE project will continue to refine the IR color measurements of the MIRION sources.

The MIRION catalog provides a sample of thousands of protoclusters across the spectrum of mass, luminosity, and age with which to explore how IR colors vary with both the physical properties and evolutionary stages of SFRs. Furthermore, roughly 300 YBs lie at distances less than 2 kpc. At this distance, instruments such as *JWST* and ALMA can study the growth of stellar embryos within protoclusters, the relationship between accretion and outflows, and the cumulative impact of outflows in protocluster environments (Section 4.2). The study of YBs continues to further our understanding of star formation in different Galactic environments and is providing opportunities for introductory astronomy students to contribute to this research (Section 4.3).

---

10 sites.google.com/view/peryscope/home

**5.2 The Power of Participatory Science**

When the editor of *Astrophysics and Space Science* approached me to write an article because of my election as a 2024 AAS Fellow, I found myself in a quandary. Although research has been one important focus of my career, I consider my most significant work to have been at the intersection of science and society, building relationships with people and organizations that represent diverse cultural and religious identities. Far and away, it is those efforts that motivated my award. Therefore, I decided to focus this article on where research and science communication meet – demonstrating how participatory research offers unique benefits to both the advancement of science, and the public perception of science and scientists.

The English translation of a famous quote by German polymath Johann Wolfgang von Goethe is, "Who wants to understand the poem must go to the land of poetry; who wishes to understand the poet must go to the poet's land" (Libquotes 2024). The meaning is clear – stepping into another's community is a powerful way to better understand that community and its constituents. Although its primary goal is to enable research that would be otherwise intractable or impossible, participatory science provides a critical benefit to society. It empowers people of all ages and from all walks of life to become collaborators with scientists, gain a better understanding of the process of science, and contribute to the pool of human knowledge.

Since 2009, *Zooniverse* has launched more than 500 hundred research projects in many subject areas across the sciences and humanities. These projects have engaged roughly 3 million volunteers, enabled more than 600 academic publications, and resulted in discoveries by citizen scientists that exceeded project expectations (Section 1.2). The MWP is a specific example of a citizen-science project that has continued to facilitate critical star-formation studies. The discovery and identification of YBs led to international media interest and the engagement of children and families (Section 2). The classroom project PERYSCOPE enables non-major introductory astronomy students to contribute to star-formation research, while learning about fundamental concepts and tools employed in astronomy (Section 4.3). PERYSCOPE fills an important niche between citizen-science projects of low complexity that enlist huge numbers of volunteers, and high-complexity research experiences available to only a few advanced students.

The *Engaging Faith-based Communities in Citizen Science through Zooniverse* initiative that was described in Section 1.2 produced models for effectively incorporating participatory science into seminary education, youth and family programs, and programs of interfaith organizations (WHT24). Through contacts I established over decades of working with faith communities, the initiative was widely communicated to audiences who may not have been attentive to other venues promoting participatory science (WHT24). Evaluation of the *Engaging* initiative demonstrated the critical importance of developing ongoing relationships between professional scientists and these communities to diversifying participation in science and improving public perceptions of science and scientists (Wolf-Chase, Hinman, & Trouille 2021).

As one of the founders of the MWP, I have been able to offer a personal perspective on the development of a citizen-science project, from its conception to the unanticipated discovery that has supplied my colleagues and me with data that is helping us elucidate differences between diverse star-forming environments. As a scientist who trod a nontraditional career path, I am able to share unique insights into engaging diverse audiences in scientific exploration. These efforts are further expounded in the Appendix. I hope my personal journey will serve to inspire early-career scientists into thinking divergently about the many ways they might combine their interests and talents to broaden the impacts of their work.

**Appendix: Reflections on a Unique Career Bridging Science and Society**

As a girl growing up in a small, all-female, suburban household during the 1960s and 1970s, science fiction (notably *Star Trek*), museums, and the space program were the principal sources of my love of science. While formal science education seemed primarily aimed at boys considering scientific careers, informal education played a large role in my decision to become a scientist. Because of this, I was strongly motivated to help inspire others who might have difficulty envisioning themselves as scientists. Therefore, after completing two traditional postdoctoral positions, I chose a very nontraditional career path.

From 1998 to 2020, I was an Astronomer at the Adler Planetarium in Chicago. Along with a few other Adler Astronomers, I simultaneously held a soft-money position at the University of Chicago, first as a Research Scientist (1998-2004), and later as a Senior Research Associate (2004-2012). This initially involved spending three days per week at the Adler Planetarium and two days per week at the University of Chicago. The theory behind the joint positions was that active researchers could "put a human face" on scientists and bring a better understanding of the process of science to the general public; however, in practice, the research and public education aspects were separated rather than integrated, and each position was eventually viewed as full time, leading to dissolution of the presence of Adler Planetarium astronomers at the University of Chicago.

While at the University of Chicago, I developed a long-term research program using the Astrophysical Research Consortium 3.5-m telescope at the Apache Point Observatory to image near-IR $H_2$ and [Fe II] spectral-line emission in regions forming high-mass ($\geq 8\ M_\odot$) and intermediate-mass (3-8 $M_\odot$) stars. This program enabled me to forge research collaborations at the University of Chicago and the Adler Planetarium, while expanding my earlier work using IR and millimeter-wave continuum and spectral-line observations, together with radiative transfer modeling, to elucidate the relationship between protostellar outflows and their birth environments (e.g., Wolf, Lada, & Bally 1990; Wolf-Chase, Walker, & Lada 1995; Wolf-Chase & Walker 1995; Wolf-Chase & Gregersen 1997; Wolf-Chase et al. 1998; O'Linger et al. 1999; Wolf-Chase, Barsony, & O'Linger 2000; Wolf-Chase et al. 2003; Barsony et al. 2010).

The near-IR observations made it possible to distinguish shocks produced by jets associated with protostellar outflows from PDRs produced by intermediate- and high-mass stars. We acquired continuum observations with the now-defunct Combined Array for Research in Millimeter-wave Astronomy (CARMA) to link jets with protostellar cores. This research provided evidence for similarities in outflows driven by protostars across a range of masses, and enabled us to place constraints on the length of the outflow phase from high-mass protostars (Wolf-Chase et al. 2012; Wolf-Chase et al. 2013; Walawender et al. 2016; Wolf-Chase et al. 2017).

At the Adler Planetarium, I served in many capacities – speaking with general audiences and writing public articles, helping to organize conferences such as the Pale Blue Dot III workshop (Meadows & Wolf-Chase 2007), and serving on teams that developed educational programs, exhibits, and sky shows. Throughout these various activities, I worked with individuals from departments across the institution who brought their diverse expertise and perspectives to each project. This provided me with valuable experience working with many talented people from outside of academia; however, my role as a "content expert" typically had little to do with my actual research. Although I was able to communicate the excitement of doing science through public interactions, for many years the research and education and public outreach (EPO) aspects of my job were mostly compartmentalized.

Over the course of time, I became acutely aware of two things: (1) the self-selected audiences that visit museums are generally not representative of the underserved, underrepresented, and/or science-apprehensive populations I was hoping to reach; and (2) communicating effectively to general audiences requires similar skills to those necessary for bridging conversations across different academic disciplines (e.g., Wolf-Chase 2004). Indeed, the more conversations to which I was privy by virtue of my unique career, the more I could identify "dialogue" that involved people talking at or past, rather than to, each other, clearly misunderstanding their conversation partners. On a public level, it was painfully apparent how misunderstandings brought on by the pervasive "culture wars" between scientific and religious understandings of the world were working to the detriment of everyone in society. This rift would not be mended by better or clearer scientific explanations, but rather through building trust and establishing long-term cross-cultural and interfaith relationships.

Astronomy, in particular, has the ability to evoke responses of awe and wonder. Hubble's Cultural Impact (2025) website tracks numerous ways that Hubble images have been showcased in popular culture. The iconic "Pillars of Creation" *Hubble* image first captured the public's imagination in 1995. Figure 11 combines *Hubble* and *JWST* views of this breathtaking SFR. Awe and wonder lie at the heart of both scientific curiosity and thoughtful religious reflection. They serve as an entry-point to building new relationships; however, building trust takes time – time that is often in short supply for scientists who hold traditional academic positions. My unique career path has provided me with the skills and time to build relationships across diverse cultures and communities of faith, and to help them engage with science in positive ways. Along the way, I have taken some lessons from the multidisciplinary field of astrobiology. I recall with great clarity how former NASA Administrator Dan Goldin addressed a meeting of the American Astronomical Society during the 1990s, asking how many astronomers had training in biology. The number was few, underscoring Goldin's point that physical and biological scientists would need to work together to address questions regarding life in the Universe. To do this, they would need to learn something about each other's "language" and methods. The first edition of the Astrobiology Primer was compiled in 2006 to assist this endeavor by helping to clarify terminology and methodology across different scientific disciplines (cf. Astrobiology Primer 1.0). This edition was superseded in 2016 and 2024 by versions 2.0 and 3.0, respectively (Astrobiology Primer 2024).

As it happens, the Editor-in-Chief of the Astrobiology Primer is astrobiologist and ordained Episcopalian priest, Rev. Dr. Lucas Mix, a member of the Society of Ordained Scientists (Society of Ordained Scientists 2024). This highlights the fact that, beyond the scientific questions, astrobiology raises ethical, philosophical, and, for many, theological, questions that are helping to establish new collaborations across the sciences and humanities. In my career, I have used questions at the intersection of astrobiology, ethics, and theology as a springboard to engage seminarians, clergy, and faith communities in science. For more than 10 years, I have written a monthly column for the e-newsletter of *The Clergy Letter Project*, an initiative to demonstrate to the public that science and religion can work together for the common good (The Clergy Letter Project Across the Cosmos / Astrobiology News Archive 2024). As an affiliated member of the Zygon Center for Religion and Science, I have taught astronomy to seminary students for 20 years. As Vice President of the Center for Advanced Study in Religion and Science (CASIRAS), an independent organization of scientists and theologians, I have helped organize webinars on themes at the intersection of science and society. I have served on advisory boards for initiatives such as the AAAS Dialogue on Science, Ethics, and Religion *Science for Seminaries* program (Science for Seminaries 2024), the University of Chicago

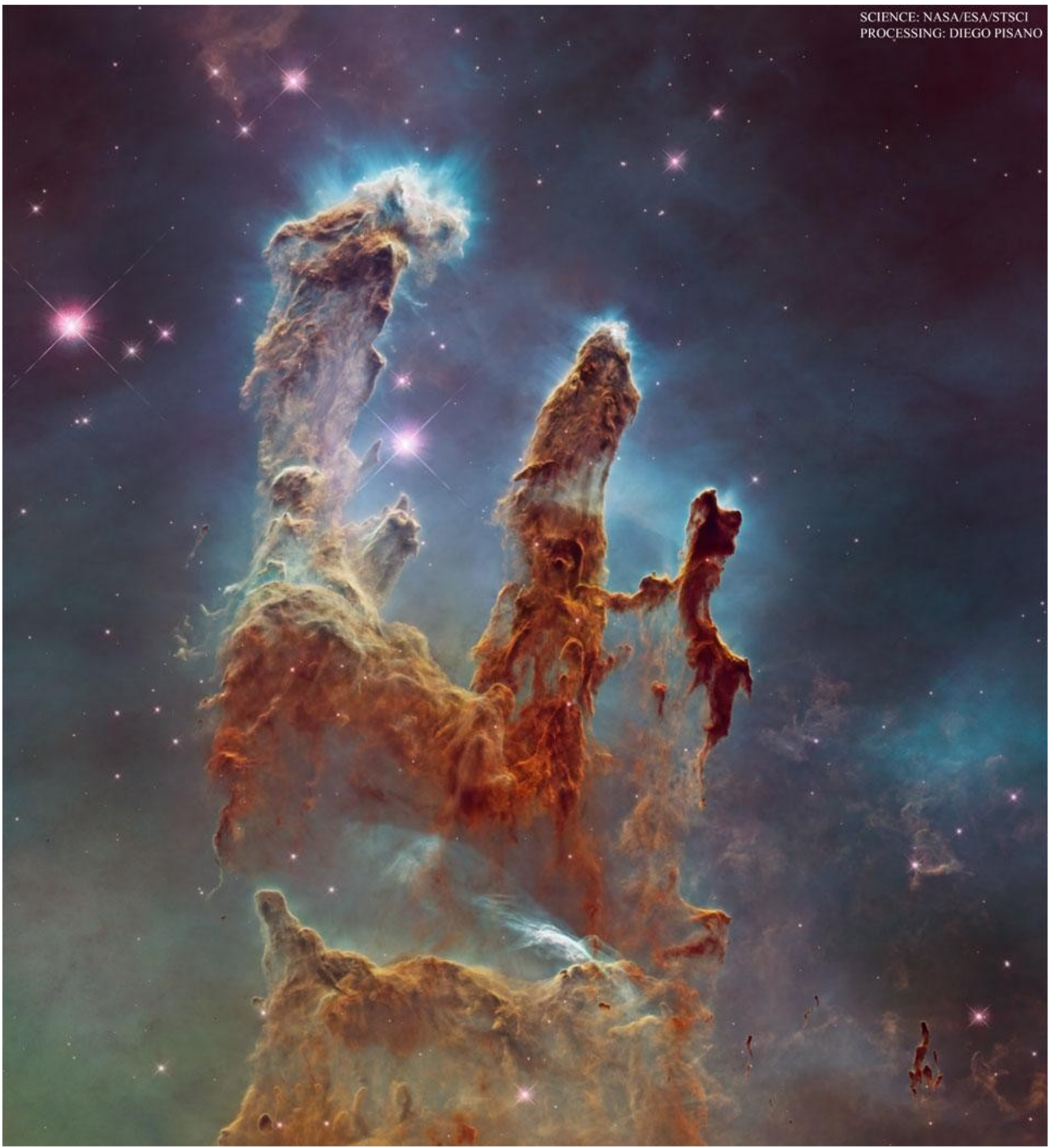

**Figure 11. M16. Pillars of Star Creation.** This image captures the dense molecular cloud material comprising the pillars in the Eagle Nebula, roughly 7000 light years distant. Featured as NASA's Astronomy Picture of the Day on October 22, 2024, this image combines visible (*Hubble*) and IR (*JWST*) light. Intense radiation from high-mass stars photo-evaporates material at the tips of the giant star-forming pillars. Image credit: NASA, ESA, CSA, STScI. Processing: Diego Pisano.

Faculty Roundtable on Science and Religion (Faculty Roundtable on Science and Religion 2024), and the Catholic Theological Union *Preaching with Sciences* project (Preaching with Sciences 2024). I also contribute to journals and books dedicated to conversations between scientists, philosophers, and theologians (e.g., Wolf-Chase 2011, 2017, 2018, 2021; WHT24).

Some of the techniques I began employing over twenty years ago to facilitate interdisciplinary discussions are now being used to catalyze new research approaches. Michigan State University philosophers have created the Toolbox Dialogue Initiative to help scientists develop collaborative skills necessary to avoid talking past each other (Toolbox Dialogue Initiative 2024). On a public level, these skills can be nurtured through long-term partnerships that involve participatory research. With upcoming astronomical observatories such as the Vera Rubin Observatory (VRO), it will be increasingly important to leverage human identifications and ML techniques to help mine the wealth of expected data; in fact, the VRO has already partnered with Zooniverse in anticipation of the roughly 5 petabytes of raw data that will be acquired during each year of operation (Vera C. Rubin Observatory Citizen Science 2024).

The biggest challenge I foresee in sustaining public engagement in participatory science is ongoing interaction with professional scientists, which is essential in maintaining cross-cultural relationships (WHT24). Virtually all scientists are limited by time and money, as are the communities we might seek to engage. There is a great need for new funding opportunities to support building scientist-community partnerships, whether these partnerships involve in-person or online research activities. Long-term partnerships could make a meaningful difference towards bridging cultural divides, to the benefit of both science and society.

**Acknowledgements**

Material presented in this paper is based upon work supported by the National Science Foundation under Grant No. 2307806; NASA under Grant No. 20-CSSFP20-001; the Murdock Charitable Trust (Grants No. NS-2016246 & SR-201811723); and the Alfred P. Sloan Foundation (Grants G-2019-12359 & G-2020-14055). Any opinions, findings, or recommendations expressed are those of the authors and do not necessarily reflect the views of the National Science Foundation, NASA, the Murdock Trust, or the Sloan Foundation. The authors want to thank the Zooniverse volunteers, former MWP leaders Rob Simpson and Matt Povich, and undergraduate students from The College of Idaho who have contributed to "yellowball" research over the years, especially Ethan Bassingwaithe, Aurora Cossairt, Bezawit Kassaye, Johanna Mori, Anupa Pouydal, Hritik Rawat, Sarah Schoultz, Makenzie Stapley, and Leonardo Trujillo. Wolf-Chase also wishes to thank Christopher Corbally for nominating her for the AAS award, and Jennifer Wiseman, Laura Trouille, and Charles Lada, for supporting the nomination. The research presented in this article utilized Astropy (Astropy Collaboration 2013; Price-Whelan et al. 2018; Price-Whelan et al. 2022), Matplotlib (Hunter 2007), NASA's Astrophysics Data System Bibliographic Services and the interactive graphical tool TOPCAT (Taylor 2005).

## Statements & Declarations

### Funding

Material presented in this paper is based upon work supported by the National Science Foundation under Grant No. 2307806; NASA under Grant No. 20-CSSFP20-001; the Murdock Charitable Trust (Grants No. NS-2016246 & SR-201811723); and the Alfred P. Sloan Foundation (Grants G-2019-12359 & G-2020-14055). Any opinions, findings, or recommendations expressed are those of the authors and do not necessarily reflect the views of the National Science Foundation, NASA, the Murdock Trust, or the Sloan Foundation.

**Competing interests:** The authors declare no competing interests.

### Author contributions

All authors have contributed to the "yellowball" research. Grace Wolf-Chase, Charles Kerton, and Kathryn Devine have been involved in all aspects of this work. Graduate student Nicholas Larose performed the distance calculations, produced Figure 6, and is leading some of the efforts described in section 4.1. Undergraduate student Maya Coleman contributed to the development of the photometry code and produced the color-color plots displayed in Figure 7. Grace Wolf-Chase has led the broader efforts to engage new audiences in participatory science. She wrote the first and revised drafts of this manuscript and all authors provided edits, comments, and approved the final manuscript.